\documentclass[showpacs, preprintnumbers,
nofootinbib, aps, prd, superscriptaddress,
10pt, showkeys, notitlepage,
twocolumn]{revtex4-1}

\usepackage{graphicx,amssymb,amsmath,amsthm,amsfonts,epsfig,natbib}
\usepackage[utf8]{inputenc}
\usepackage[linktocpage,breaklinks]{hyperref}
\usepackage[usenames,dvipsnames]{color}
\usepackage{tikzsymbols}
\usepackage{epstopdf}
\usepackage{aas_macros}
\usepackage{tensor}
\usepackage{mathtools}
\usepackage{amsbsy}
\usepackage{bm}

\definecolor{darkred}{rgb}{0.5,0,0}
\definecolor{darkgreen}{rgb}{0,0.5,0}
\definecolor{darkblue}{rgb}{0,0,0.5}
\definecolor{prussian}{rgb}{0.0, 0.19, 0.33}

\hypersetup{colorlinks=true, citecolor=darkblue, linkcolor=darkblue,
urlcolor = darkblue}

\newcommand{\be}{\begin{equation}}
\newcommand{\ee}{\end{equation}}
\newcommand{\bear}{\begin{eqnarray}}
\newcommand{\eear}{\end{eqnarray}}
\newcommand{\ba}{\[\begin{aligned}}
\newcommand{\ea}{\end{aligned}\]}

\newcommand{\p}{\prime}
\newcommand{\pp}{\prime\prime}

\newcommand{\cD}{{\cal D}}

\newcommand{\cC}{{\cal C}}

\newcommand{\cU}{{\cal U}}

\newcommand{\rK}{{\rm K}}

\newcommand{\rph}{{\rm ph}}

\begin{document}
\title{Post-Kerr black hole spectroscopy}

\begin{abstract}
  One of the central goals of the newborn field of gravitational wave
  astronomy is to test gravity in the highly nonlinear, strong field
  regime characterizing the spacetime of black holes. In particular,
  ``black hole spectroscopy'' (the observation and identification of
  black hole quasinormal mode frequencies in the gravitational wave
  signal) is expected to become one of the main tools for probing the
  structure and dynamics of Kerr black holes. In this paper we take a
  significant step towards that goal by constructing a ``post-Kerr''
  quasinormal mode formalism. The formalism incorporates a
  parametrized but general perturbative deviation from the Kerr metric
  and exploits the well-established connection between the properties
  of the spacetime's circular null geodesics and the fundamental
  quasinormal mode to provide approximate, eikonal limit formulae for
  the modes' complex frequencies. The resulting algebraic toolkit can be
  used in waveform templates for ringing black holes with the purpose
  of measuring deviations from the Kerr metric.  As a first
  illustrative application of our framework, we consider the
  Johannsen-Psaltis deformed Kerr metric and compute the resulting
  deviation in the quasinormal mode frequency relative to the known
  Kerr result.
\end{abstract}

\author{Kostas Glampedakis}
\email{kostas@um.es}
\affiliation{Departamento de F\'isica, Universidad de Murcia, Murcia, E-30100, Spain}
\affiliation{Theoretical Astrophysics, University of T\"ubingen, Auf der Morgenstelle 10, T\"ubingen, D-72076, Germany}

\author{George Pappas}
\email{georgios.pappas@tecnico.ulisboa.pt}
\affiliation{Department of Physics and Astronomy,
The University of Mississippi, University, MS 38677, USA}
\affiliation{Departamento de F\'isica, CENTRA, Instituto Superior T\'ecnico,
Universidade de Lisboa, Avenida Rovisco Pais 1, 1049 Lisboa, Portugal}

\author{Hector O. Silva}
\email{hosilva@phy.olemiss.edu}
\affiliation{Department of Physics and Astronomy,
The University of Mississippi, University, MS 38677, USA}

\author{Emanuele Berti}
\email{eberti@olemiss.edu}
\affiliation{Department of Physics and Astronomy,
The University of Mississippi, University, MS 38677, USA}
\affiliation{Departamento de F\'isica, CENTRA, Instituto Superior T\'ecnico,
Universidade de Lisboa, Avenida Rovisco Pais 1, 1049 Lisboa, Portugal}

\date{{\today}}

\maketitle

%%%%%%%%%%%%%%%%%%%%%%%%%%%%%%%%%%%%%%%%%%%%%%%%%%%%

\section{Introduction}
\label{sec:intro}

The direct observation of merging black hole binaries during the first
observation run (O1) of Advanced LIGO marked a milestone in the
history of astronomy and fundamental physics.  The three detections
(GW150914~\cite{Abbott:2016blz}, GW151226~\cite{Abbott:2016nmj} and
GW170104~\cite{Abbott:2017vtc}), plus a fourth candidate LVT151012
that is likely of astrophysical
origin~\cite{TheLIGOScientific:2016pea}, provide a formidable
laboratory to test general relativity (GR) in the strong-gravity
regime~\citep{Yunes:2013dva,Berti:2015itd,TheLIGOScientific:2016src}. More
detections are expected in the near future.

The first event, GW150914, was particularly striking for its high
signal-to-noise ratio (SNR), and because it allowed a direct observation of
the strong-field merger and ringdown of the binary. Similar
observations in the future may allow us to do ``black hole
spectroscopy'': as first proposed by
Detweiler~\cite{Detweiler:1980gk}, the measurement of multiple
oscillation frequencies and damping times of the merger remnant may
identify Kerr black holes, just like atomic lines allow us to identify
atomic elements~\cite{Dreyer:2003bv,Berti:2005ys}. However, given our
current understanding of astrophysical black hole formation, the
detection of several modes will require either more advanced detectors
on Earth and in
space~\cite{Berti:2007zu,Berti:2016lat,Belczynski:2016ieo} or better
data analysis techniques~\cite{Yang:2017zxs}.

Vishveshwara discovered quasinormal modes (QNMs) via time evolutions
in the Schwarzschild spacetime~\cite{Vishveshwara:1970zz}. Soon
afterwards, Press computed QNM frequencies in a short-wavelength
(eikonal) approximation~\cite{Press:1971wr}, and Goebel (inspired by
Ames and Thorne's study of collapsing
stars~\cite{1968ApJ...151..659A}) understood that there is an intimate
relation between QNM frequencies and unstable null
geodesics~\cite{Goebel:1972} (see
e.g.~\cite{Kokkotas:1999bd,Berti:2009kk} for reviews). The imaginary
part of the modes is similarly related to the Lyapunov exponent, (the
inverse of) the instability timescale associated with null geodesic
motion~\cite{Cardoso:2008bp}.

This connection between null geodesics and QNMs has been
explored in depth for Kerr black
holes~\cite{Ferrari:1984zz,Dolan:2010wr,Yang:2012he,Yang:2012pj,Yang:2013uba}. Our
goal here is to extend this connection beyond the Kerr spacetime, and
to turn it into a practical scheme to test experimentally whether a
set of QNM frequencies (such as those potentially observable by LIGO)
is consistent with the dynamics of the Kerr spacetime.

The remainder of the paper is structured as follows. In
Section~\ref{sec:scheme} we briefly discuss the inherent difficulty in
computing QNMs in non-GR theories of gravity and motivate the use of the
eikonal limit approximation. Section~\ref{sec:summary} provides a practical
summary of the post-Kerr toolkit and the main result of the paper, namely,
the eikonal QNM formulae. In Section~\ref{sec:photon_gen} we study circular
null geodesics (``light rings'') in a general stationary axisymmetric
spacetime. In Sec.~\ref{sec:PKphotonring} we specialize these results to
the case of a general post-Kerr metric and calculate the associated light
ring frequency and Lyapunov exponent. In Section~\ref{sec:applications}
we consider the Johannsen-Psaltis (JP) deformed Kerr metric and compute eikonal
QNM frequencies for both small and generic deviations from Kerr.
Our concluding remarks can be found in Section~\ref{sec:conclusions}.
Some technical material and lengthy equations are collected in the
Appendices.

%%%%%%%%%%%%%%%%%%%%%%%%%%%%%%%%%%%

\subsection{The eikonal post-Kerr parametrization scheme}
\label{sec:scheme}

The post-Kerr scheme of this paper is based on the use of eikonal
limit formulae for a QNM's real and imaginary parts. This approach is
dictated more by necessity than by choice. Computing QNMs in generic
non-GR theories is unrealistic, because black hole perturbation theory
should be developed (in principle) for any given choice of the field
equations.  There have been attempts to build such a
formalism for specific classes of theories, such as Horndeski
gravity. However these attempts are usually limited to spherical
symmetry, and they often lead to the conclusion that large classes of
black hole solutions are unstable~\cite{Kobayashi:2012kh,Kobayashi:2014wsa}.
As far as we know, QNM frequencies in modified gravity were
computed only in a handful of cases, specifically in
Einstein-dilaton-Gauss-Bonnet~\cite{Blazquez-Salcedo:2015ets,Blazquez-Salcedo:2016enn,Blazquez-Salcedo:2017txk}
and dynamical Chern-Simons~\cite{Cardoso:2009pk,Molina:2010fb} theories, and
even then only for spherically symmetric black hole solutions.
These calculations are therefore of limited utility in data analysis
applications, both because they must be developed on a case-by-case
basis, and because the remnant of a binary black hole merger is almost
inevitably a rotating black hole\footnote{Producing a Schwarzschild
remnant requires an astrophysically unrealistic fine-tuning of the
parameters of the merging binary, such that the individual black
hole spins exactly cancel the orbital angular momentum at
merger~\cite{Buonanno:2007sv,Berti:2007nw}.}.

Relying on the eikonal limit/QNM link is a reasonable
alternative strategy, reinforced by the fact that it is known to
perform surprisingly well in the case of the Kerr spacetime as long as
one is interested in approximating the \emph{fundamental} QNM for a
given $(\ell,\,m)$
multipole~\cite{Dolan:2010wr,Yang:2012he,Yang:2012pj,Yang:2013uba}. This
is the mode associated with the spacetime's circular null
geodesics and with the peak of the radial potential that determines the
properties of wave scattering after separating angular variables in the
perturbation equations (see e.g.~\cite{Berti:2014bla}).

The light ring/QNM correspondence should be broadly valid in
modified theories of gravity that can be used as tests of GR
provided that (i) gravitational waves propagate with the speed of light
(e.g. Lorentz-violating theories likely fall short of this
requirement~\cite{Berti:2015itd}) and (ii) deviations from the Einstein
field equations (and deviations of the corresponding black hole solutions
from Kerr) can be parametrized by some small perturbative
parameter~\cite{Yagi:2016jml}.

Our post-Kerr formalism implicitly assumes a  ``Kerr-like''
situation, in the sense that the non-Kerr spacetime should admit a
single geodesic light ring structure that can be physically connected
to the observed QNM signal. In fact, these fundamental QNMs are known
to dominate the spacetime's perturbative dynamics as it happens, for
example, in the case of general relativistic Kerr black holes and
ultracompact stars~\cite{Kokkotas:1999bd,Berti:2009kk}.

This restriction aside, the post-Kerr scheme can handle equally well
``bumpy Kerr metrics" (i.e. makeshift deformed Kerr metrics that are
not consistent solutions of any gravitational field equations, see
e.g.~\cite{Johannsen:2013rqa,Yagi:2016jml} for reviews) and known black hole spacetime
solutions produced by modified theories of gravity (but for which the
QNM perturbation calculation is often very
complicated or impractical)~\cite{Berti:2015itd,Yagi:2016jml}.

As an illustrative application, in this paper we study the JP ``bumpy
Kerr'' metric~\cite{Johannsen:2011dh} (see
Section~\ref{sec:applications} below). There is an abundance of
``bumpy'' black hole metrics that could be considered for data
analysis applications, such as those proposed in
Refs.~\cite{Cardoso:2014rha,Konoplya:2016jvv,Younsi:2016azx}.

Besides focusing on the fundamental QNMs, in this paper we exclusively
study the $\ell=|m|$ angular multipoles. There is good reason for this
choice, since these modes are considered to be the most powerful
emitters of gravitational waves, and as a consequence the most easily
detectable by gravitational wave
observatories~\cite{Berti:2005ys,Buonanno:2006ui,Berti:2007fi,Berti:2007zu,Kamaretsos:2011um,Gossan:2011ha,
  Kamaretsos:2012bs,Agathos:2013upa,London:2014cma,Meidam:2014jpa,Berti:2016lat,Cardoso:2016ryw,Yang:2017zxs}.
At the same time they are the easiest to model with the eikonal
approximation, since they are associated with equatorial photon orbits
(more specifically, a positive/negative $m$ corresponds to
prograde/retrograde orbital motion).

In order to facilitate the comparison between Kerr and non-Kerr QNMs
we need to express the former in an eikonal form. To this end we
introduce the ``offset'' function  $\beta_\rK(a)$ defined by
\be
\omega_\rK = \sigma_\rK + \beta_\rK,
\label{eq:beta}
\ee
where $\omega_\rK$ is the exact Kerr QNM frequency and $\sigma_\rK$ is
the analytically known, eikonal-limit
formula~\cite{Ferrari:1984zz,Mashhoon:1985cya}.
The offset function $\beta_\rK (a)$ can be obtained via numerical fits
to tabulated Kerr QNM data~\cite{Berti:2005ys,Berti:2009kk}. These
fits and their accuracy are discussed in Appendix~\ref{app:offset}.

An eikonal QNM frequency $\sigma$ can be obtained from the properties
of the equatorial light ring of a given non-Kerr spacetime. Then, an
\emph{observed} QNM frequency $\omega_{\rm obs}$, gleaned from
gravitational wave data, is match-filtered by the complex-valued
``template'',
\be
\omega_{\rm obs} = \sigma + \beta_\rK.
\ee
A genuine Kerr QNM signal obviously implies $\sigma = \sigma_\rK$.  On
the other hand, the combination of a non-Kerr spacetime \emph{and} a
non-Kerr light ring structure is bound to lead to a mismatch
\be
\omega_{\rm obs}  - \omega_\rK = \sigma -\sigma_\rK \neq 0.
\ee
In practice (and taking into account the recent gravitational wave
observations of merging black holes) we would expect to face situations
where the deviation from Kerr is small. This means that it makes sense
to employ a simpler post-Kerr form $\sigma = \sigma_\rK + \delta \sigma $
and get
\be
\delta \sigma = \omega_{\rm obs} - \omega_\rK,
\ee
with $\delta \sigma $ encoding the deviation from the Kerr metric. A
large portion of this paper is devoted to the explicit calculation of
this parameter; the final outcome will be a fully \emph{algebraic}
expression in terms of $M$, $a$ and leading-order metric deviations from
Kerr evaluated at the Kerr light ring.
The derivation of a similar algebraic result for a general non-Kerr
spacetime is \emph{not} possible, for the simple reason that the radial
location of the light ring comes as a solution of a transcendental
equation.

The proposed parametrization is a simple \emph{null test}:
$ \delta \sigma =0$ if and only if the spacetime is exactly described
by the Kerr metric. This scheme fails in the special (and presumably
highly unlikely!)  case of a non-Kerr metric with a Kerr light
ring. It is also obvious that, if present, the measured deviation from
Kerr will carry some amount of inaccuracy due to the use of the Kerr
offset $\beta_\rK$.

%%%%%%%%%%%%%%%%%%%%%%%%%%%%%%%%%%%

\subsection{The post-Kerr QNM toolkit summarized}
\label{sec:summary}

This section collects the key elements of the post-Kerr formalism
in the form of a ``toolkit'' that can be used in the construction
of parametrized  QNM templates. The detailed calculations leading
to these results are presented in subsequent sections. A remark
about notation: the label ``ph'' identifies Kerr parameters evaluated
at the Kerr circular photon orbit $r_\rph$ while Kerr functions at an
arbitrary radius are labelled by a ``K''. Non-Kerr parameters are
identified by a subscript ``0''.

The main idea is to work with a simple, perturbative post-Kerr metric
correction $h_{\mu\nu}$, such that a general axisymmetric-stationary metric
is expressed in the form
\be
g_{\mu\nu} = g_{\mu\nu}^\rK (r) + \epsilon h_{\mu\nu} (r) + {\cal O}(\epsilon^2),
\ee
where $g_{\mu\nu}^\rK$ is the Kerr metric and we only keep
leading-order terms in the perturbative parameter $\epsilon$.
Also, the $\theta$-dependence has been suppressed, as we are
considering equatorial orbits.

This expansion can be used to find modifications to the Kerr light
ring radius (the upper/lower sign corresponds to prograde/retrograde motion)
\be
r_\rph = 2M \left \{ 1 + \cos \left [ \frac{2}{3} \cos^{-1} \left (\mp \frac{a}{M} \right ) \right ] \right \},
\label{rph_book}
\ee
and to the Kerr light ring angular frequency
\be
\Omega_\rph = \pm \frac{M^{1/2}}{r^{3/2}_\rph \pm a M^{1/2}}.
\label{OmKbook}
\ee
The result is
\begin{align}
&r_0 = r_\rph + \epsilon \delta r  + {\cal O} (\epsilon^2), \\
&\Omega_0 =    \Omega_\rph + \epsilon \delta \Omega_0 + {\cal O} (\epsilon^2),
\end{align}
where the shifts $\delta r$ and $\delta \Omega_0$ can be computed
by expanding the light ring equation. The explicit forms of these
post-Kerr modifications are
\begin{align}
\delta r  &=  -\frac{1}{6} h^\p_{\varphi\varphi} +  \frac{(r_\rph -M)^{-1}}{6 r_\rph}
\Big\{\,  C_{tt} h_{tt}^\p   \pm 4\left ( C_{t\varphi}  h_{t\varphi}^\p    \right.
\nonumber \\
& \left.  +\, D_{t\varphi} h_{t\varphi}  \right ) + 4 M \left[  (3r^2_\rph +a^2) h_{tt}  + h_{\varphi\varphi} \right]  \, \Big\}
\label{dr_shift}
\end{align}
 and
\begin{align}
\delta \Omega_0   & =  \mp \left ( \frac{M}{r_\rph} \right )^{1/2} \Big [\,  h_{\varphi\varphi}
\pm   \left (\frac{r_\rph}{M} \right )^{1/2}  ( r_\rph + 3M)  h_{t\varphi}
\nonumber \\
&  +  ( 3 r_\rph^2 +a^2) h_{tt}  \, \Big ]/ \left [(r_\rph -M) ( 3r^2_\rph +a^2) \right ],
\label{dOm_shift}
\end{align}
where
\begin{align}
D_{t\varphi} & =    (M r_\rph)^{1/2} ( r_\rph + 3M),
\\
C_{tt} &=  -( a^2 + 63 M^2) r_\rph^2 +  ( 135 M^2 -11 a^2) M r_\rph
\nonumber \\
& -60 M^2 a^2,
\\
C_{t\varphi} &=   ( Mr_\rph )^{1/2} ( 3 M r_\rph -2r_\rph^2 -a^2 ).
\end{align}
In these expressions a prime stands for $d/dr$ and $h_{\mu\nu}$ and
its derivatives are to be evaluated at $r_\rph$.

Apart from the light ring frequency shift, the formalism makes
contact with the local divergence rate of photon orbits grazing
the light ring. These orbits can be approximated near the light
ring as,
\be
r(t) \approx r_0 \left (\, 1 + \cC e^{\pm \gamma_0 t} \, \right ),
\ee
where $\cC$ is a constant. The divergence rate of photon orbits
grazing the light ring $\gamma_0$ (which is essentially the
Lyapunov exponent for these orbits) is also modified with respect
to its Kerr value:
\be
\gamma_0 = \gamma_\rph + \epsilon \delta \gamma_0 + {\cal O} (\epsilon^2).
\ee
The Kerr expression for this parameter is~\cite{Ferrari:1984zz,Mashhoon:1985cya}
\be
\gamma_\rph = 2\sqrt{3M}  \frac{\Delta_\rph \Omega_\rph}{r_\rph^{3/2} (r_\rph -M)},
\ee
where  $\Delta_\rph = r^2_\rph -2M r_\rph + a^2$.

For the post-Kerr shift we find the rather complicated result
%\begin{widetext}
\begin{align}
\delta \gamma_0 &=\mp  \frac{4M^2}{\sqrt{3}}  \left\{\, (r_\rph + 3M ) \left[\, G_{tt} h_{tt}^{\pp}
+  G_{\varphi\varphi} h_{\varphi\varphi}^{\pp}  +  2 Z_{tt} h^\p_{tt}   \right. \right.
\nonumber \\
& \left. \left .   +  2 Z_{\varphi\varphi} h^\p_{\varphi\varphi}
\pm (M / r_\rph )^{1/2} \left ( \,  G_{t\varphi} h_{t\varphi}^{\pp} + 4 Z_{t\varphi} h_{t\varphi}^\p  \right ) \,  \right. \right.
\nonumber \\
& \left. \left.   +  6 E_{rr} h_{rr}   \, \right]  + 2M \left ( S_{tt} h_{tt}  + S_{\varphi\varphi} h_{\varphi\varphi}  \right. \right.
\nonumber \\
& \left. \left. \pm S_{t\varphi} h_{t\varphi} \right ) \, \right\} /  [  \Delta_\rph r_\rph^5 (r_\rph +3M)^{2} (r_\rph -M)^{3}  ],
\label{dgam0_pK}
\end{align}
%\end{widetext}
where the various coefficients are listed in Appendix~\ref{sec:coeffs}.

The eikonal-limit formulae for the  QNM frequency
$\sigma = \sigma_{\rm R} + i \sigma_{\rm I}$ associated with the
light ring are
\be
\sigma_{\rm R} = m \Omega_0, \qquad \sigma_{\rm I} = - \frac{1}{2} | \gamma_0 |.
\label{eik0}
\ee
Their post-Kerr approximation is the principal result of this paper:
\begin{align}
\sigma_{\rm R} & = m \left ( \Omega_\rph + \epsilon \delta \Omega_0 \right ),
\label{eik_R}
\\
\sigma_{\rm I} & =  - \frac{1}{2} | \gamma_\rph + \epsilon \delta \gamma_0 |.
\label{eik_Im}
\end{align}
Both quantities are functions of the Kerr parameters $M,a$ and of
the post-Kerr metric correction $h_{\mu\nu}$ evaluated at the Kerr light ring
$r_\rph$. The imaginary part $\sigma_{\rm I}$ in addition depends on the first
and second derivatives of $h_{\mu\nu}$.

%%%%%%%%%%%%%%%%%%%%%%%%%%%%%%%%%%%%%%%%%%%%%%%%%%%

\section{Light ring in a general stationary axisymmetric spacetime}
\label{sec:photon_gen}

In this section we consider  circular photon orbits in a spacetime
that is stationary and axisymmetric, but otherwise arbitrary.
The special case of the Kerr metric is textbook material that
we review (mostly to establish notation) in Appendix~\ref{sec:Kcirc}.

\subsection{Equatorial photon orbits}

As pointed out earlier, we are interested in equatorial null geodesics.
The four-velocity normalization gives,
\be
g_{tt} (u^t)^2 + 2 g_{t\varphi} u^t u^\varphi + g_{rr} (u^r)^2 + g_{\varphi\varphi} (u^\varphi)^2 = 0.
\label{element1}
\ee
Given the imposed symmetries, any geodesic has a conserved
energy $E= -u_t$ and angular momentum $L= u_\varphi $ (both per unit mass).
%given by
%\begin{align}
%E &=  -u_t = -g_{tt} u^t - g_{t\varphi} u^\varphi,
%\\
%L &= u_\varphi   = g_{\varphi\varphi} u^\varphi + g_{t\varphi} u^t.
%\end{align}
These relations can be inverted:
\begin{align}
u^t &= \frac{1}{\cD} \left (\, g_{\varphi\varphi} E + g_{t\varphi} L \, \right ),
\\
u^\varphi &= -\frac{1}{\cD} \left ( \, g_{tt} L + g_{t\varphi} E \,  \right ),
\label{uph_sol}
\end{align}
where $\cD \equiv g_{t\varphi}^2 - g_{tt} g_{\varphi\varphi}$.
From these we can immediately obtain the (coordinate) azimuthal angular frequency,
\be
\Omega  =  \frac{d\varphi}{dt} =  \frac{u^\varphi}{u^t}
= -\frac{g_{tt} L + g_{t\varphi} E }{g_{t\varphi} L + g_{\varphi\varphi} E }.
\label{Omdef}
\ee
After eliminating the two velocities in (\ref{element1}) we obtain
an effective potential equation for the radial motion:
\be
\cD g_{rr} ( u^r )^2 = g_{\varphi\varphi} E^2 +2 g_{t\varphi} EL + g_{tt} L^2 \equiv  V_{\rm eff}.
\label{Veff0}
\ee
If the orbit has a turning point ($u^r=0$) at $r_p$, then
\be
 g_{\varphi\varphi} (r_p)  +2 g_{t\varphi} (r_p) b + g_{tt} (r_p) b^2 = 0,
\ee
where we introduced the impact parameter:
\be
\label{bdef}
b \equiv \frac{L}{E}.
\ee
At the same time, Eq.~(\ref{element1}) can be written as
\be
g_{tt} (r_p) + 2 g_{t\varphi} (r_p) \Omega_p + g_{\varphi\varphi} (r_p) \Omega^2_p = 0,
\label{Omeq1}
\ee
where $\Omega_p \equiv \Omega (r_p)$.
Thus, at any turning point
\be
\Omega_p = \frac{1}{b} \quad \Leftrightarrow \quad E = \Omega_p L.
\label{bOm}
\ee
Obviously, this simple relation will hold for a circular
photon orbit as well.

%%%%%%%%%%%%%%%%%%%%%%%%%%%%%%%%%%%%%%%%%%

\subsection{Light ring}
\label{sec:ring}

Circular motion at a radius $r=r_0$ must meet the following
two conditions:
\be
\label{eq:circular}
V_{\rm eff} (r_0)=0, \qquad V^\p_{\rm eff} (r_0) = 0.
\ee
Both equations lead to quadratics of $b$:
\bear
&& g_{tt} b^2  +2 g_{t\varphi} b +  g_{\varphi\varphi}    = 0,
\label{Veff1}
\\
\nonumber \\
&& g_{tt}^\p b^2 +2 g_{t\varphi}^\p b  +  g_{\varphi\varphi}^\p   = 0.
\label{dVeff1}
\eear
In solving these we follow the same steps as in the
corresponding analysis of Kerr orbits (see Appendix~\ref{sec:Kcirc}).
From Eq.~(\ref{dVeff1}) we get
\be
b = \frac{1}{g_{tt}^\p} \left (\, - g_{t\varphi}^\p \mp  W^{1/2} \,\right ), \quad
W =  (g^\p_{t\varphi})^2 - g_{tt}^\p g_{\varphi\varphi}^\p,
\label{bsol1}
\ee
where the upper (lower) sign corresponds to prograde
(retrograde) motion. Inserting this in Eq.~(\ref{Veff1}) we
obtain the light ring equation:
\begin{multline}
g_{\varphi\varphi}  (g^\p_{tt} )^2  + 2 g_{tt} (g^\p_{t\varphi} )^2 -g^\p_{tt} \left (\, g_{tt} g_{\varphi\varphi}^\p
+ 2 g_{t\varphi} g_{t\varphi}^\p   \,\right )
\\
\mp 2 W^{1/2} \left  (\, g_{t\varphi} g^\p_{tt} - g_{tt} g^\prime_{t\varphi} \, \right )    = 0.
\label{phring}
\end{multline}
The angular frequency $\Omega_0$ at the light ring\footnote{It
should be pointed out that the two angular frequency expressions
(\ref{bOm}) and (\ref{Om0_sol}) hold for orbits of massive particles
as well. Interestingly, the latter expression has a hidden ``symmetry''
that allows it  to take  the equivalent ``inverted'' form $\Omega_0 =
( - g_{t\varphi}^\p  \pm W^{1/2}) /g_{\varphi\varphi}^\p$. This is
of course a consequence of the high symmetry of the underlying
spacetime.} is obtained with the help of Eq.~(\ref{bOm}),
\be
\Omega_0 =\frac{ g_{tt}^\p}{ - g_{t\varphi}^\p  \mp W^{1/2}  }.
\label{Om0_sol}
\ee
In the Kerr metric limit, $g_{\mu\nu} \to g_{\mu\nu}^\rK$,
Eqs.~(\ref{bsol1})-(\ref{Om0_sol}) reduce to well known expressions
[cf. Eqs.~(\ref{bKsol2}), (\ref{rphK}) and (\ref{OmKbook2}) in
Appendix~\ref{sec:Kcirc}].

%%%%%%%%%%%%%%%%%%%%%%%%%%%%%%%%%%%%%%%%%%%%%%%%%%%%

\subsection{Orbiting near the light ring}
\label{sec:near_ring}

The association between the light ring structure and the
spacetime's fundamental QNM frequency requires, apart from
the properties of the circular photon orbits themselves, the
study of orbits that approach the light ring from far away and
asymptotically tend to become circular. In other words these
are parabolic-like orbits with their periapsis located at $r_0$.
The rate with which these orbits ``zoom-whirl" towards the
light ring is the key parameter connected to the imaginary part
of the eikonal QNM (in the Kerr spacetime there is also a direct
link between this parameter and the curvature of the wave potential
at the location of its maximum).

Considering non-circular equatorial photon orbits, we follow the
textbook approach and use the auxiliary radial variable $\cU=1/r$.
Then,
\be
\frac{d\cU}{d\varphi}  = - \cU^2 \frac{u^r}{u^\varphi}.
\ee
After eliminating $u^\varphi$ and $u^r$ with the help of
(\ref{uph_sol}) and (\ref{Veff0}), we arrive at a Binet-type
equation describing the shape $r(\varphi)$ of the orbit:
\be
\left ( \frac{d\cU}{d\varphi} \right )^2 = \frac{\cU^4 \cD}{g_{rr}}
\frac{\left (\, g_{tt} b^2 + 2 g_{t\varphi} b + g_{\varphi\varphi}  \,\right )}{( g_{tt} b + g_{t\varphi} )^2}
\equiv f(\cU).
\label{binet_gen}
\ee
Given that  $\cU_0=1/r_0$ is a turning point, we should have
\be
f(\cU_0) = f^\prime (\cU_0 ) = 0 = \frac{d f}{d\cU} (\cU_0).
\ee
The portion of the orbit near the light ring can be studied via an
expansion
\be
\cU = \cU_0 + \varepsilon \cU_1 + {\cal O} (\varepsilon^2), \qquad \varepsilon \ll 1.
\ee
The leading order perturbative term solves
\be
\frac{d\cU_1}{d\varphi}  = \pm  \kappa_0 \, \cU_1,
\label{binet1}
\ee
where we have defined
\be
\kappa^2_0 = \frac{1}{2} \frac{d^2 f}{d\cU^2} (\cU_0) =  \frac{f^{\prime\prime} (\cU_0)}{2 \cU_0^4} .
\ee
For  the second $r$-derivative of $f$ at $\cU_0$ we find
\be
f^{\pp} (\cU_0) =  \cU^4_0 \frac{ \cD}{g_{rr}}
\frac{\left (\, g_{tt}^{\pp} b^2 + 2 g_{t\varphi}^{\pp} b
+ g_{\varphi\varphi}^{\pp}  \,\right )}{( g_{tt} b + g_{t\varphi} )^2},
\ee
and this leads to
\be
\kappa^2_0 =  \frac{\cD \left (\, g_{tt}^{\pp} b^2 + 2 g_{t\varphi}^{\pp} b
+ g_{\varphi\varphi}^{\pp}  \,\right )}{2g_{rr}( g_{tt} b + g_{t\varphi} )^2}.
\label{kappa_gen}
\ee
Eq.~(\ref{binet1}) admits the exponential solutions
\be
\cU_1 = C e^{\pm \kappa_0 \varphi}  \qquad (C = \mbox{const}).
\ee
This can be written as a  time domain expression with the
simple substitution $\varphi = \Omega_0 t + \mbox{const}$. The
resulting equation describes the convergence/divergence of our
light ring-grazing orbits as a function of time:
\be
\cU (t) \approx \cU_0 + \varepsilon C e^{\pm \gamma_0 t},
\ee
where $C$ has been rescaled and
\be
\gamma_0 \equiv | \kappa_0 \Omega_0 |.
\label{gamma_gen}
\ee
This $\cU(t)$ expression illustrates the role of the parameter
$\gamma_0$ as a measure of the local divergence rate of null
geodesics at $r_0$. In other words, $\gamma_0$ is the Lyapunov
exponent of these orbits.

%%%%%%%%%%%%%%%%%%%%%%%%%%%%%%%%

\section{Post-Kerr light ring formalism and eikonal QNM}
\label{sec:PKphotonring}

So far our analysis has been based on the use of a general
stationary-axisymmetric metric.  As we have seen in the preceding
sections, we can derive the light ring's angular frequency
$\Omega_0$ [Eq.~(\ref{Om0_sol})] and Lyapunov exponent $\gamma_0$
[Eq.~(\ref{gamma_gen})] as functions of the metric $g_{\mu\nu}$ and
its derivatives evaluated at the light ring's radius $r_0$. Once these
parameters have been calculated, the eikonal QNM frequency can be
obtained immediately via Eqs.~(\ref{eik0}).
The main drawback of this general approach is that $r_0$ is not
known beforehand, but must be computed by solving  Eq.~(\ref{phring})
which, in general, is a transcendental expression.

%%%%%%%%%%%%%%%%%%

\subsection{Post-Kerr light ring}
\label{sec:pKring}

A more practical approach, closer to the spirit of producing
QNM templates that could be used as a measuring device of the
``Kerrness'' of black holes seen by gravitational wave detectors,
is that of adopting a simpler post-Kerr metric of the form
\be
g_{\mu\nu} = g_{\mu\nu}^\rK (r) + \epsilon h_{\mu\nu} (r) + {\cal O}(\epsilon^2),
\ee
and working to first order with respect to the metric deviation
$h_{\mu\nu}$. Note that we use an index K to label Kerr spacetime
parameters and we only consider the equatorial hypersurface.

According to this recipe, the orbital frequency (\ref{Om0_sol})
can be approximated as
\be
\Omega (r) = \Omega_\rK (r) + \epsilon \delta \Omega (r) + {\cal O} (\epsilon^2),
\ee
where
\begin{align}
\Omega_\rK (r) &= \pm \frac{M^{1/2}}{r^{3/2} \pm aM^{1/2}},
\\
\delta \Omega (r)  &=  \mp  \frac{1}{4} \Omega_\rK   \left (\frac{r}{M} \right)^{1/2}   \Big [\,
 2 h_{t\varphi}^\p +   \Omega_\rK h^\p_{\varphi\varphi}
\nonumber \\
& +\,  \frac{\Omega_\rK}{M}  \left (\, r^3  \pm 2aM^{1/2} r^{3/2} + Ma^2 \, \right )   h_{tt}^\p  \, \Big ].
\end{align}
These expressions need to be combined with the modified light ring radius
\be
r_0 = r_\rph + \epsilon \delta r  + {\cal O} (\epsilon^2),
\ee
where $r_\rph$ is the Kerr light ring [see Eq.~(\ref{rph_book})]. The
shift $\delta r$ can be computed by expanding the light ring equation
(\ref{phring}). After some algebra and repeated use of the Kerr light
ring equation (\ref{rphK}) we find:
\begin{align}
\delta r  &=  -\frac{1}{6} h^\prime_{\varphi\varphi} + \frac{ (r_\rph -M)^{-1}}{6 r_\rph}
\Big\{\,  C_{tt} h_{tt}^\prime   \pm 4 \left ( C_{t\varphi}  h_{t\varphi}^\prime   \right.
\nonumber \\
& \left. \quad  +\, 4 D_{t\varphi} h_{t\varphi} \right ) + 4M \left[  (3r^2_\rph +a^2) h_{tt}   + h_{\varphi\varphi} \right]  \, \Big\},
\label{dr_sol}
\end{align}
where from now on it is understood that $h_{\mu\nu}$ and its derivatives
are to be evaluated at $r_\rph$. The coefficients $C_{tt}, D_{t\varphi},
C_{t\varphi} $ have already been given in Section~\ref{sec:summary}.

The angular frequency at the light ring is given by the expansion,
\begin{align}
\Omega_0 &= \Omega_\rph + \epsilon \left [ \,\delta \Omega_\rph + B_\rph \delta r  \,\right ]
+ {\cal O} (\epsilon^2)
\nonumber \\
& \equiv    \Omega_\rph + \epsilon \delta \Omega_0,
\label{OmpK1}
\end{align}
where $\Omega_\rph = \Omega_\rK (r_\rph)$ is the Kerr light
ring frequency,  and $\delta\Omega_\rph = \delta \Omega (r_\rph)$.
The net frequency shift $\delta \Omega_0 $ receives contributions
from both $\delta \Omega$ and $\delta r$. For these partial
contributions we find
\begin{align}
 \delta \Omega_\rph & =  \mp \frac{ (M/ r_\rph)^{1/2}}{( r_\rph + 3M )^{2}} \Big [\, h_{\varphi\varphi}^\p
 +  (3 r_\rph^2 + a^2) h_{tt}^\p
 \nonumber \\
& \quad  \pm  \left (\frac{r_\rph}{M} \right )^{1/2} ( r_\rph + 3M )  h_{t\varphi}^\p  \, \Big ],
\\
B_\rph &= \mp \, 6  \frac{ (M/ r_\rph)^{1/2}}{( r_\rph + 3M )^{2}}.
\end{align}
After assembling the two pieces we obtain the total post-Kerr frequency shift,
\begin{align}
\delta \Omega_0   & =  \mp \left ( \frac{M}{r_\rph} \right )^{1/2} \Big [\,  h_{\varphi\varphi}
\pm   \left (\frac{r_\rph}{M} \right )^{1/2}  ( r_\rph + 3M)  h_{t\varphi}
\nonumber \\
&  +\,  ( 3 r_\rph^2 +a^2) h_{tt}  \, \Big ]/ [(r_\rph -M) ( 3r^2_\rph +a^2) ].
\label{dOm_tot}
\end{align}
Interestingly, this expression turns out to be independent of
the metric derivatives $h^\p_{tt}, h^\p_{t\varphi}, h^\p_{\varphi\varphi}$.

Having obtained the post-Kerr light ring radius and frequency,
we next turn our attention to photon ring-grazing orbits and the
associated Lyapunov exponent.

%%%%%%%%%%%%%%%%%%%%%%%%%%%%%%

\subsection{Post-Kerr Lyapunov exponent}
\label{sec:pKgrazing}

In this section we derive a post-Kerr formula for the
Lyapunov exponent $\gamma_0 = |\kappa_0 \Omega_0|$, see
Eq.~(\ref{gamma_gen}). To this end, and given that we already
have a post-Kerr expression for $\Omega_0$, we only need to
expand the $\kappa_0$ parameter. From (\ref{kappa_gen}) we find
\be
\kappa^2_0 = \kappa^2_\rph + \epsilon \left (\, \kappa^2_{\rm \delta r}  + \kappa^2_h \, \right )
+ {\cal O} (\epsilon^2).
\label{kappa_pK1}
\ee
The first term is the Kerr $\kappa_\rK^2 (r)$ evaluated at $r=r_\rph$:
\be
\kappa^2_\rph =  \frac{12 M \Delta_\rph^2}{r^3_\rph (r_\rph -M)^2}.
\ee
The term $\kappa_{\delta r}$ originates from $\kappa_\rK (r)$
when evaluated at the post-Kerr light ring
$r_0 = r_\rph + \epsilon \delta r $. We find,
\be
\kappa^2_{\delta r} = -\frac{24 M R_\rph \delta r}{r^4_\rph  (r_\rph -M)^3} \left (\frac{M}{r_\rph} \right )^{3/2},
\ee
where
\begin{align}
R_\rph &= \left (19 M^2 + 26 a^2 \right ) M r^2_\rph + 3 M a^2 \left (8 M^2 + 7a^2\right)
\nonumber \\
&  \quad  - \left( 54 M^4   + 40 M^2 a^2  -4a^4  \right ) r_\rph.
\end{align}
Finally, the term $\kappa_h^2$ is produced by the $h_{\mu\nu}$ perturbation:
\be
\kappa^2_h =  - \frac{4 \Delta_\rph H_\rph}{r^4_\rph (r_\rph -M)^3} \left ( \frac{M}{r_\rph} \right )^{3/2}.
\label{kappa_h}
\ee
The quantity $H_\rph$ is an algebraic function of $h_{\mu\nu}$
and its first and second order derivatives:
\begin{align}
& H_\rph =   \frac{1}{2} \left ( \frac{r_\rph}{M} \right )^{1/2} (r_\rph -M) \left [\, 6 \Delta_\rph^2 h_{rr}
-r^2_\rph   \Delta_\rph h^{\prime\prime}_{\varphi\varphi}  \right.
\nonumber \\
& \left. -\, 6r_\rph (r_\rph-2M) h_{\varphi\varphi}  \, \right ]   \pm r_\rph \Delta_\rph   W_{t\varphi} h_{t\varphi}^\p
\nonumber \\
& \mp\, 2 r_\rph M_{t \varphi} \left [\,  r_\rph \Delta_\rph h^{\prime\prime}_{t\varphi} + 6(r_\rph-2M) h_{t\varphi} \, \right ]
\nonumber \\
&
+  \left ( \frac{r_\rph}{M} \right )^{1/2}  r_\rph \left [\,  3K_{tt}  h_{tt}  - \Delta_\rph \left (  Q_{tt} h_{tt}^\p + r_\rph J_{tt} h_{tt}^{\pp} \right ) \, \right ]
\nonumber \\
& +  r_\rph \Delta_\rph  \left ( \frac{r_\rph}{M} \right )^{1/2} (2r_\rph -5M)  h_{\varphi\varphi}^\prime,
\label{Hph_eq}
\end{align}
where the coefficients $ Q_{tt}, J_{tt}, K_{tt}, W_{t\varphi},
M_{t\varphi}$ are binomials in $r_\rph$, see Appendix~\ref{sec:coeffs}.
The total post-Kerr $\kappa_0$ is:
\be
\kappa_0 = \kappa_\rph  + \epsilon \frac{\kappa^2_{\delta r} + \kappa^2_h}{2\kappa_\rph}
\equiv \kappa_\rph + \epsilon \delta \kappa_0.
\ee
With the help of our previous results this leads to
\be
\delta \kappa_0 = -\frac{2 M}{\sqrt{3}}  \frac{N_\rph}{r_\rph^5 \Delta_\rph}  (r_\rph -M)^{-3},
\label{dkappa0}
\ee
where $N_\rph$ takes the symbolic form
\begin{align}
 N_\rph & =  \left ( M r_\rph \right )^{1/2} \left( \, G_{\varphi\varphi} h_{\varphi\varphi}^{\pp} + G_{tt} h_{tt}^{\pp}
+ 2 Z_{tt}  h_{tt}^\p +  2 Z_{\varphi\varphi}  h_{\varphi\varphi}^\p \right.
\nonumber \\
&\left. \quad + \,  2 E_{tt} h_{tt}   + 2 E_{\varphi\varphi} h_{\varphi\varphi}   +  6 E_{rr} h_{rr}   \, \right)
\nonumber \\
&  \quad \pm M \left ( G_{t\varphi} h_{t\varphi}^{\pp}  + 4  Z_{t\varphi} h_{t\varphi}^\p + 8 E_{t\varphi} h_{t\varphi} \right ).
\label{Nph}
\end{align}
All of the coefficients appearing in this expression are
binomials with respect to $r_\rph$ and are listed in
Appendix~\ref{sec:coeffs}.

The post-Kerr expanded $\gamma_0$ takes the form,
\be
\gamma_0 = \kappa_\rph \Omega_\rph  + \epsilon \left (\, \Omega_\rph  \delta \kappa_0
+ \kappa_\rph \delta \Omega_0 \, \right ) \equiv \gamma_\rph + \epsilon \delta \gamma_0.
\ee
After assembling the previous results we obtain
\begin{widetext}
\begin{align}
\label{dgamma0}
\delta \gamma_0 &=\mp \frac{4M^2}{\sqrt{3} \Delta_\rph r_\rph^5} (r_\rph + 3M)^{-2} (r_\rph -M)^{-3}
\Big [\, (r_\rph + 3M ) \left(\, G_{tt} h_{tt}^{\pp} + G_{\varphi\varphi} h_{\varphi\varphi}^{\pp}
+  2 Z_{tt} h^\p_{tt}    +  2 Z_{\varphi\varphi} h^\p_{\varphi\varphi} + 6 E_{rr} h_{rr}    \, \right)
\nonumber \\
&  \pm \, (r_\rph +3M) \left ( \frac{M}{ r_\rph } \right )^{1/2} \left(\,  G_{t\varphi} h_{t\varphi}^{\pp} + 4 Z_{t\varphi} h_{t\varphi}^\p  \right)
 + 2M \left ( S_{tt} h_{tt} + S_{\varphi\varphi} h_{\varphi\varphi}  \pm S_{t\varphi} h_{t\varphi}  \right )\, \Big ],
\end{align}
\end{widetext}
where $S_{tt}, S_{t\varphi}, S_{\varphi\varphi}$ can also be
found in Appendix~\ref{sec:coeffs}.

Having at hand the post-Kerr deviations $\delta \Omega_0$
[Eq.~(\ref{dOm_tot})] and $\delta \gamma_0$ [Eq.~(\ref{dgamma0})]
for the light ring orbital frequency and Lyapunov exponent,
it is straightforward to proceed to our ultimate goal: the
construction of the post-Kerr QNM eikonal formulae. These final
results have already been presented in Section~\ref{sec:summary}
[Eqs.~(\ref{eik_R}) and (\ref{eik_Im})].

%%%%%%%%%%%%%%%%%%%%%%%%%%%%%%%%%%%%%%
\section{A post-Kerr application: the Johannsen-Psaltis metric}
\label{sec:applications}

As an example of a non-Kerr spacetime we now consider the JP ``bumpy
Kerr'' metric. In the JP model~\cite{Johannsen:2011dh}, the ``bumps''
are introduced by the function:
\be
h (r,\theta) = \sum_{k=0}^{\infty}\left( \varepsilon_{2k}
+  \varepsilon_{2k+1}\frac{M r}{\Sigma}\right)\left(\frac{M^2}{\Sigma}\right)^k,
\ee
where $\Sigma = r^2 + a^2 \cos^2\theta$ and $\varepsilon_k$ are freely adjustable
parameters. Johannsen and Psaltis showed
that the first two parameters $\varepsilon_0$ and $\varepsilon_1$ must
be zero if we require asymptotic flatness, and that experimental
constraints imply that $\varepsilon_2$ must be small:
$|\varepsilon_2|<4.6\times10^{-4}$~\cite{Johannsen:2011dh}. For
these reasons we can parametrize perturbations of the Kerr metric
using a single function
\be
\label{JPdeviation}
h (r,\theta) =  \varepsilon_3\frac{M^3 r}{\Sigma^2}
\ee
that is proportional to the first unconstrained parameter,
$\varepsilon_3$.
The modified metric components read
\begin{align}
& g_{tt}^{\rm JP} = (1+ h) g^\rK_{tt}, \quad g_{t\varphi}^{\rm JP} = (1+ h) g^\rK_{t\varphi},
\\
& g_{rr}^{\rm JP} = g_{rr}^\rK (1+h) \left (1+h \frac{a^2 \sin^2\theta}{\Delta} \right )^{-1},
\\
%\nonumber \\
& g_{\varphi\varphi}^{\rm JP} = g_{\varphi\varphi}^\rK  + h a^2 \left (1 + \frac{2Mr}{\Sigma} \right ) \sin^4\theta,
\quad g_{\theta\theta}^{\rm JP} = g_{\theta\theta}^\rK,
\end{align}
where
$\Delta=r^2-2Mr+a^2$.  When viewed as a truncated equatorial post-Kerr
metric, $g_{\mu\nu}^{\rm JP} = g_{\mu\nu}^\rK + \epsilon h_{\mu\nu}^{\rm JP} + {\cal
  O}(\epsilon^2)$, the relevant JP metric components are
\begin{align}
& h_{tt}^{\rm JP}   =  - \left (1-\frac{2M}{r} \right ) \left (\frac{M}{r} \right )^3,
\\
& h_{t\varphi}^{\rm JP}  =  -\frac{2Ma}{r} \left (\frac{M}{r} \right )^3,
\\
& h_{rr}^{\rm JP}  =  \frac{r^4}{\Delta^2} \left (1-\frac{2M}{r} \right )  \left (\frac{M}{r} \right )^3,
\\
& h_{\varphi\varphi}^{\rm JP} =   a^2  \left (\frac{M}{r} \right )^3  \left (1 + \frac{2M}{r} \right ),
\\
& h_{\theta\theta}^{\rm JP}   =0.
\end{align}
The deformation parameter $\varepsilon_3$ is generally assumed to take
values up to $\mathcal{O}(10)$~\cite{Maselli:2016nuu}, and from
the asymptotic structure of the metric it would correspond to a GR
quadrupole deformation of the form
\be
Q_{\textrm{JP}} = \left[ -(a/M)^2+\varepsilon_3 \right] M^3
= Q_{\textrm{Kerr}} + \varepsilon_3 M^3.
\ee
Strictly speaking, the JP metric is not a vacuum spacetime, therefore the moments do not enter as simple coefficients in the metric,
as one would have in the vacuum exterior of an object in GR. Therefore the statement above should be taken with a grain of salt.

%%%%%% FIG. 1 %%%%%%%%%%%%%
\begin{figure}[b]
\includegraphics[width=0.5\textwidth]{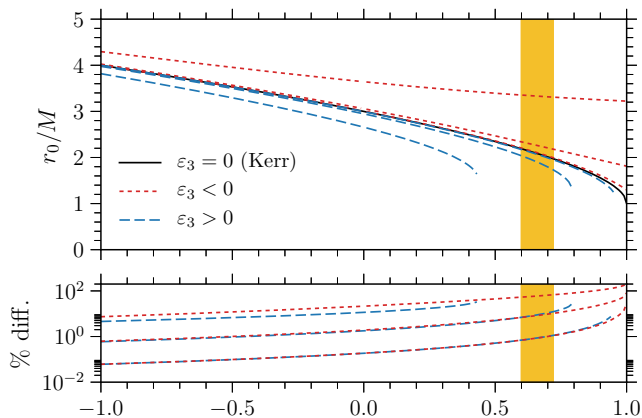}
\caption{Radius $r_0$ of the JP light ring. The solid black line
  corresponds to the Kerr light ring.  Dashed (dotted) curves show the
  deviation from the Kerr spacetime (solid line) for $\varepsilon_3>0$
  ($\varepsilon_3<0$) when we set $|\varepsilon_3|=0.1,\,1,\,10$ in
  Eq.~(\ref{JPdeviation}) ($10$ being when the deviations are the largest).
  The lower panel shows the relative difference
  ($\equiv 100 \times |(y_{\rm K} - y_{\rm JP})/y_{\rm K}|$) between the
  radius of the JP and Kerr light rings.
  }
\label{fig:nonKerr_exactPh}
\end{figure}

%%%%%%%% FIG. 2 %%%%%%%%%%
\begin{figure*}[t]
\includegraphics[width=0.49\textwidth]{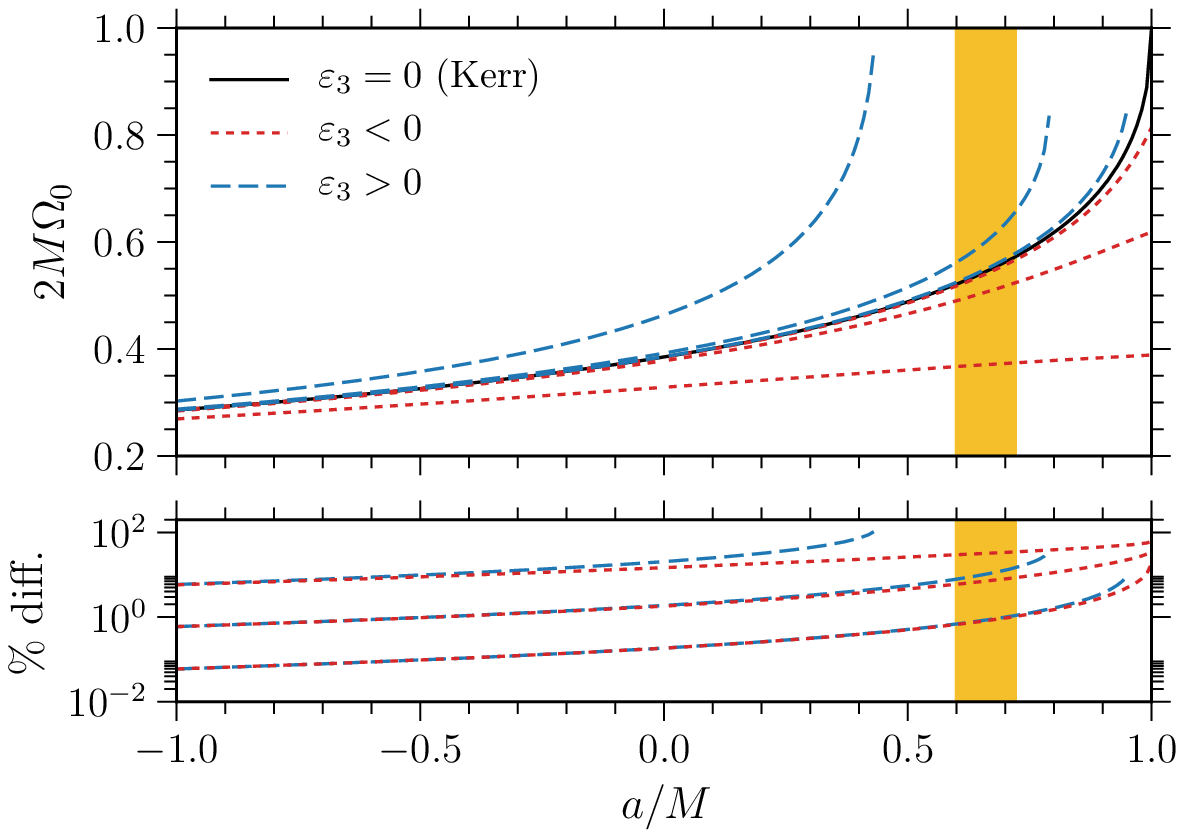}
\includegraphics[width=0.49\textwidth]{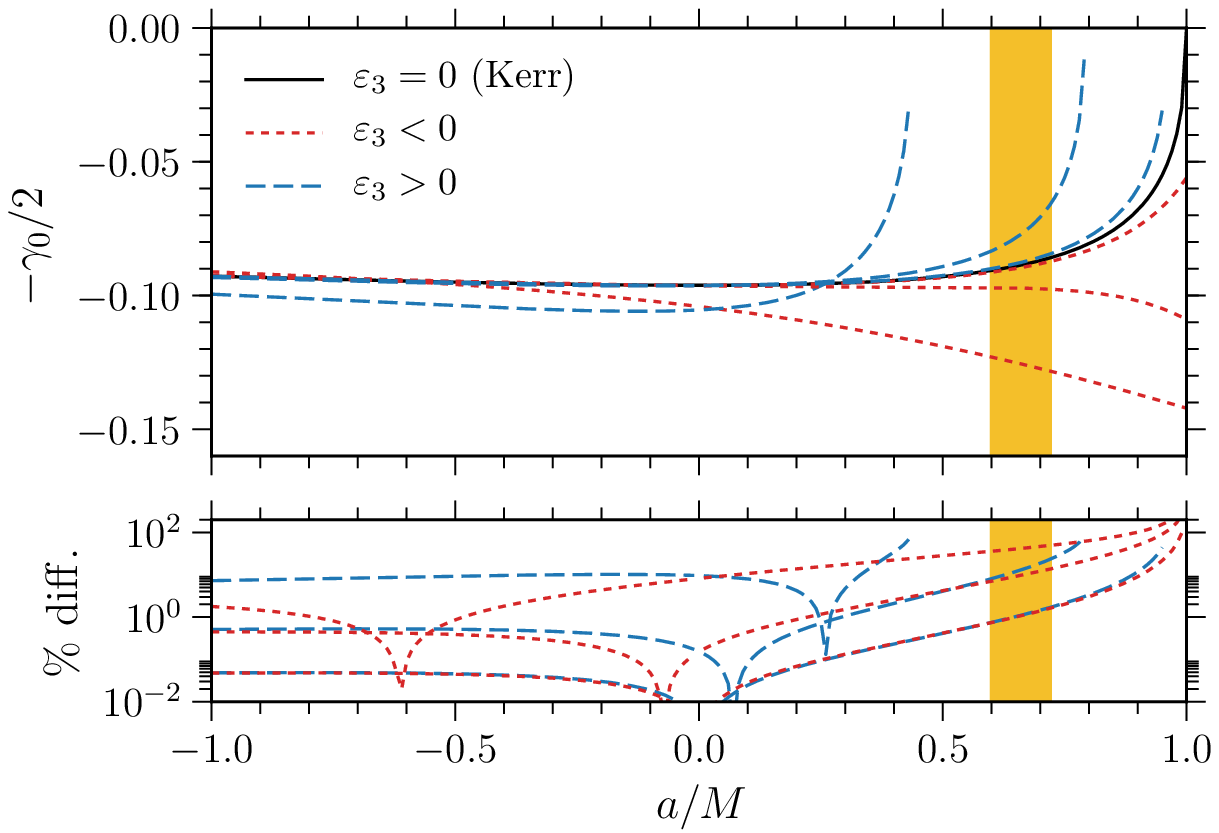}
\caption{Top-left: real part of the fundamental JP QNM frequency for $\ell=|m|=2$.
Top-right: imaginary part of the fundamental JP QNM frequency
$\ell=|m|=2$. The solid curve corresponds to the Kerr QNM frequency while
the deviations induced by $\varepsilon_3 \neq 0$ are shown for
$\varepsilon_3=\pm0.1$ (curves closest to the Kerr result), $\pm1$ and $\pm10$
(curves that deviate the most from the Kerr result). QNM frequencies
corresponding to positive (negative) values of $\varepsilon_3$ are shown in
dashed (dotted) lines.
Lower-left and lower-right: the relative difference
($\equiv 100 \times |(y_{\rm K} - y_{\rm JP})/y_{\rm K}|$) on
$2 M \Omega_0$ and $-\gamma_0/2$ for the JP metric with respect to the
Kerr metric.  The (yellow) shaded band marks GW150914's measured spin value
$a= 0.67^{+0.05}_{-0.07}M$~\cite{Abbott:2016blz}.
}
\label{fig:nonKerr_exact}
\end{figure*}

%%%%%%%%%%%%%%%%%%%%%%%%%%%%

\subsection{Numerical calculation of the light ring}

To determine the circular photon orbit we need to solve
Eqs.~(\ref{eq:circular}). For the JP metric, these reduce to the
system
\begin{align}
0&= ( \varepsilon _3 M^3+4 r^3) (a^2-b^2)+6 M r^2 (a-b)^2
+6 r^5,
\label{ringJP0}
\\
0&= \left(\varepsilon _3 M^3+r^3\right) \left[2 M (a-b)^2+r (a^2-b^2)\right]+r^6.
\label{ringJP}
\end{align}
Unfortunately this system does not admit a simple analytic solution,
but we can find the radius $r_0$ and impact parameter $b_0$ of the
circular photon orbit numerically. In Fig.~\ref{fig:nonKerr_exactPh}
we compare the radius of the Kerr light ring against the
corresponding radius $r_0$ for the JP metric with selected values of
the parameter $\varepsilon_3$ that correspond to either oblate
($\varepsilon_3<0$) or prolate ($\varepsilon_3>0$) deformations. For
concreteness we set $|\varepsilon_3| = 0.1$ (curves that barely
deviate from the Kerr curve), $|\varepsilon_3| = 1$ and
$|\varepsilon_3| = 10$ (curves for which the deviation from Kerr is
the largest).

In the left panel of Fig.~\ref{fig:nonKerr_exact} we plot the QNM
frequency $2M\Omega_0$ obtained using the JP light ring frequency
$\Omega_0=1/b_0$ for selected values of the parameter $\varepsilon_3$,
and we compare it to the corresponding frequency computed using the
Kerr light ring frequency $\Omega_\rph$.

The Lyapunov exponent \eqref{gamma_gen}
for the JP non-Kerr spacetime, after some algebra, takes the form
\be
\gamma_0=\gamma_\rK (r_0,b_0) \left [ 1 + \varepsilon_3 \left ( \frac{M}{r_0} \right)^3  f_1
+ \varepsilon_3^2 \left ( \frac{ M}{r_0} \right )^6 f_2 \right ]^{1/2},
\label{JPLyapunovExact}
\ee
where
\begin{align}
f_1 &= \frac{r_0 \left(b_0^2-a^2\right) +2 M (a-b_0)^2-4 r_0^3}{4 M (a-b_0)^2+r_0 (a^2-b_0^2)},
\\
f_2 & = \frac{4 a b_0 M-2 a^2 (M+r_0)+2 b_0^2 (r_0-M)+5 r_0^3}{4 M (a-b_0)^2+r_0 (a^2-b_0^2)},
\end{align}
and
\begin{align}
& \gamma_\rK(r,b) =-\frac{\sqrt{3}}{b}  r^{-13/2} \left [ 4 M (a-b)^2+r (a^2-b^2) \right ]^{1/2}
\nonumber  \\
 & \times \left [2 M (a-b)^2+r (a^2-b^2) \right]   \left [2 a M+b (r-2 M) \right ]
\end{align}
is a function that formally gives the Kerr Lyapunov exponent as
$\gamma_\rph=\gamma_\rK (r_\rph,b_\rph)$, where $(r_\rph,b_\rph = 1/ \Omega_\rph )$
are given by Eqs.~(\ref{rph_book}) and (\ref{OmKbook}).
In the right panel of Fig.~\ref{fig:nonKerr_exact} we show the
imaginary part of the fundamental $\ell=m=2$ QNM obtained from the
Lyapunov exponent of the JP metric with different values of the
parameter $\varepsilon_3$.

%%%%%%%% FIG. 3 %%%%%%%%%%
\begin{figure*}[t]
\includegraphics[width=0.98\textwidth]{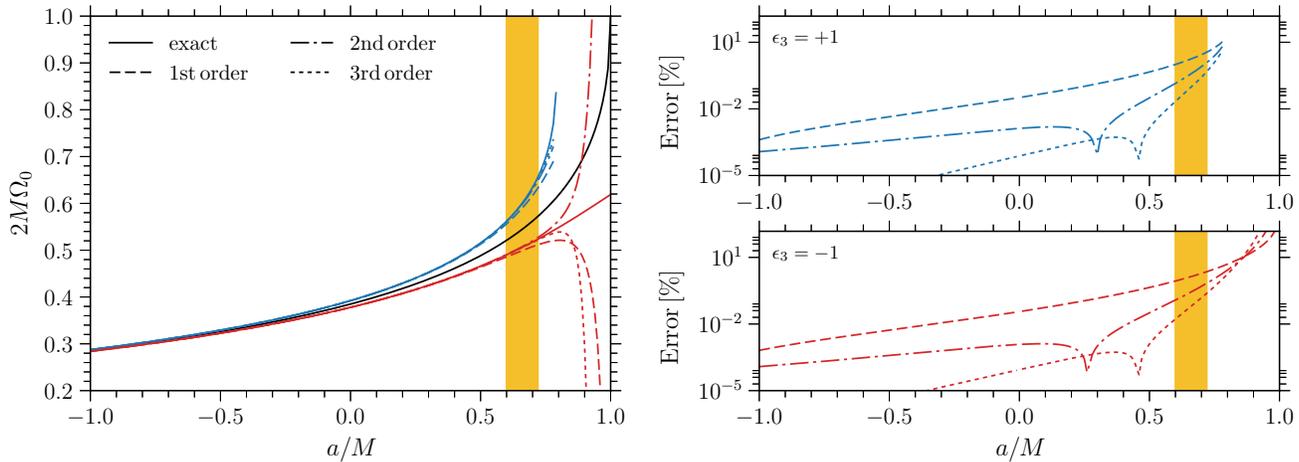}
\caption{
Comparison of the approximate real part of the QNM
frequencies $\Omega_0=1/b_0$, where $b_0$ is given by Eq.~(\ref{b0})
and reexpanded to the relevant order in $\varepsilon_3$,
against the exact JP result for $\varepsilon_3=\pm1$. In all panels
the solid lines correspond to the exact results, the dashed lines to
the first-order approximation, the dot-dashed to the second-order
approximation and the dotted to the third-order approximation. The
blue (red) lines corresponds to $\varepsilon_3 = +1$ ($-1$). Left:
Comparison between the exact results against the perturbative
calculation.  Observe that the convergence of the perturbative
expansion is slow for $a / M \gtrsim 0.7$.
Top-right: the percent error
$(\equiv 100\times|(y_{\rm approx} - y_{\rm exact})/y_{\rm exact}|)$ for
$\epsilon_3 = +1$.
Bottom-right: the percent error for $\epsilon_3 = -1$. As in
Fig.~\ref{fig:nonKerr_exact}, the (yellow) shaded band
marks GW150914's measured spin value
$a= 0.67^{+0.05}_{-0.07}M$~\cite{Abbott:2016blz}.
}
\label{fig:nonKerr_Re_approx}
\end{figure*}

%%%%%%% FIG. 4 %%%%%%%%%%%
\begin{figure*}[htb]
\includegraphics[width=0.973\textwidth]{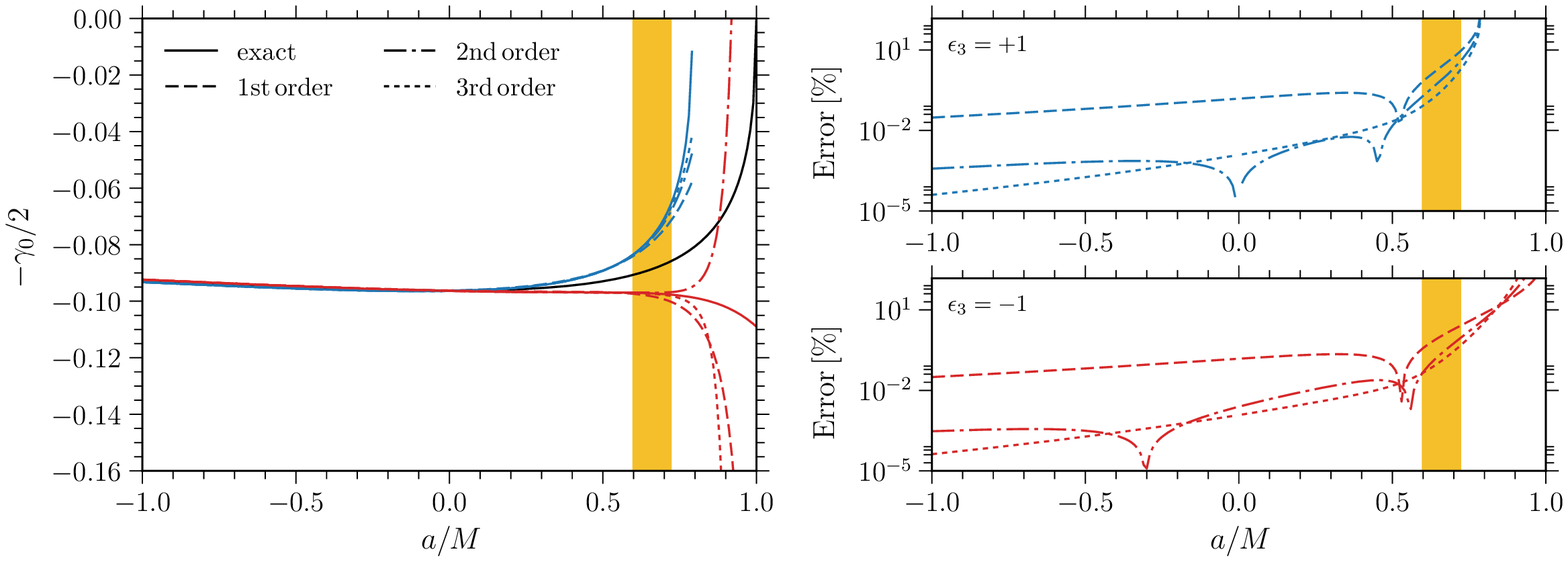}
\caption{
Comparison of the approximate imaginary part of QNMs
(\ref{JPLyapunovApprox}) against the exact non-Kerr result
(\ref{JPLyapunovExact}) for $\varepsilon_3=\pm1$. In all panels the
solid lines correspond to the exact results, the dashed lines to the
first-order approximation, the dot-dashed to the second-order
approximation and the dotted to the third-order approximation.
The blue (red) lines corresponds to $\varepsilon_3 = +1$ ($-1$). Left:
Comparison between the exact results against the perturbative
calculation.  Observe that the convergence of the perturbative
expansion is slow for $a / M \gtrsim 0.7$.
Top-right: the percent error
$(\equiv 100\times|(y_{\rm approx} - y_{\rm exact})/y_{\rm exact}|)$ for
$\epsilon_3 = +1$.
Bottom-right: the percent error for $\epsilon_3 = -1$. As in
Fig.~\ref{fig:nonKerr_exact}, the (yellow) shaded region
marks GW150914's measured spin value
$a= 0.67^{+0.05}_{-0.07}M$~\cite{Abbott:2016blz}.
}
\label{fig:nonKerr_Im_approx}
\end{figure*}

%%%%%%%%%%%%%%%%%%%%%%%%%%%%%
\subsection{Approximate solution for the light ring}

Instead of solving the system of Eqs.~(\ref{ringJP0}) and
(\ref{ringJP}) numerically, we can look for approximate solutions
assuming small perturbations around the Kerr metric, and considering
$\varepsilon_3$ as the expansion parameter. Then we can write a series
expansion for the light ring radius and for the impact parameter:
\begin{align}
&r_0=r_\rph+ \delta r_1 \varepsilon_3  + \delta r_2
  \varepsilon_3^2  + \delta r_3 \varepsilon_3^3  +\ldots \,,
\\
\label{b0}
&b_0=b_\rph+ \delta b_1 \varepsilon_3  + \delta b_2
  \varepsilon_3^2  + \delta b_3 \varepsilon_3^3  +\ldots \,.
\end{align}
Assuming this ansatz, the system can be solved order by order in
$\varepsilon_3$. The first few coefficients obtained in this way
($\delta r_i$ and $\delta b_i$ with $i=1,\,2,\,3$) are listed in
Appendix~\ref{app:JPcoeffs}.

The photon ring  frequency and Lyapunov exponent can be expanded
in a similar way with respect to $\varepsilon_3$:
\begin{align}
\Omega_0 &= \frac{1}{b_\rph}  -   \frac{2 \delta b_1}{b_\rph^2} \varepsilon_3
- \frac{2 }{b_\rph^3} \left (b_\rph \delta b_2 - \delta b_1^2 \right )  \varepsilon_3^2
+ {\cal O}\left (\varepsilon_3^3 \right ),
\label{Om0_JP_approx}
\\
\gamma_0 &=\gamma_\rph+ \delta\gamma_1 \varepsilon_3
+ \delta\gamma_2 \varepsilon^2_3 +  {\cal O}\left (\varepsilon_3^3 \right ).
\label{JPLyapunovApprox}
\end{align}
The leading-order coefficient $\delta\gamma_1$ is listed in
Appendix~\ref{app:JPcoeffs}. We omit higher-order coefficients because
they are lengthy and unenlightening.
As a sanity check, we have verified that for $h_{\mu\nu} = h^{\rm JP}_{ \mu\nu}$
the general post-Kerr results, Eqs.~(\ref{dOm_shift}) and (\ref{dgam0_pK}),
exactly match the ${\cal O} (\varepsilon_3) $ precision JP expressions (\ref{Om0_JP_approx})
and (\ref{JPLyapunovApprox}).

In Figs.~\ref{fig:nonKerr_Re_approx} and~\ref{fig:nonKerr_Im_approx}, we show
the accuracy of this perturbative scheme when used to calculate the real
(associated with $b_0$) and the imaginary (associated
with  $\gamma_0$) parts of the $\ell = |m| = 2$ QNM
frequency for $\varepsilon_3 = \pm 1$.
We see that the convergence is rather slow
for $a/M \gtrsim 0.8$, although the errors with respect to the exact
calculation are typically small otherwise.

\section{Concluding remarks}
\label{sec:conclusions}

As described in the preceding sections, the construction of
eikonal limit formulae for the fundamental $\ell=|m|$ QNMs of
a general post-Kerr spacetime is a straightforward procedure,
although the final expressions unavoidably involve some algebraic
complexity. The main results of the paper, Eqs.~(\ref{eik_R})
and (\ref{eik_Im}), can be used to produce QNM spectra for any
stationary axisymmetric metric that can be written as a
perturbation of Kerr. Our illustrative case study of the JP
spacetime and the comparison against the ``exact" results one
can obtain with that metric has helped us to gauge the accuracy
of the linear approximation (on top of that related to the use of
the eikonal/geodesic approximation).

The validity of our strategy to establish a null test according to the
recipe laid out in Section~\ref{sec:scheme} is also confirmed by
numerical simulations, which show that fundamental QNMs with
$\ell=m=2,\,3,\,4$ should dominate the ringdown signal in comparable
mass black hole
mergers~\cite{Buonanno:2006ui,Berti:2007fi,London:2014cma,Bhagwat:2016ntk}.
However, in its complete form, the black hole spectroscopy program
will require the inclusion of nonequatorial modes (i.e. $|m| < \ell$),
enabling it to handle spinning mergers, where at least the
$\ell=2,\, m=1$ multipole is known to be excited to a significant
level\footnote{The relative contribution of the asymmetric
  $(\ell,m) = (2,\,1)$ multipole with respect to the first few
  $\ell=m$ modes is a function of the spins of the merging black
  holes~\cite{Kamaretsos:2011um,Kamaretsos:2012bs}.  For small (or
  exactly zero) spins this multipole is comparable to the $(4,\,4)$
  mode and much below the $(3,\,3)$ mode. This arrangement can be
  drastically altered in rapidly spinning systems and for certain spin
  orientations, with the (2,\,1) multipole even becoming comparable to
  the dominant quadrupole (2,\,2).}.
Within our framework, this extension calls for the study of
nonequatorial circular photon orbits in non-Kerr spacetimes, and
it is an important goal earmarked for follow-up work. In that respect,
a great deal of progress in relating nonequatorial photon orbits with
$|m| < \ell$ QNMs has already been achieved in the context of Kerr
black holes \cite{Dolan:2010wr,Yang:2012he}.

Another key topic that ought to be addressed by future work
is the actual detectability and data analysis of QNM signals.
Black hole spectroscopy, as a probe for testing the Kerr metric,
relies on the extraction of more than one QNM frequency/multipole
from the observed gravitational wave signal. This exciting prospect
would require a much louder QNM signal (typically a factor $\sim 10$
boost in the SNR) than those thus far detected by
LIGO~\cite{Berti:2007zu,Berti:2016lat,Belczynski:2016ieo}.
Moreover, very recent work \cite{Thrane:2017ll} suggests that
the intrinsic precision of spectroscopy could be affected by the
uncertainty in the transition from the merger's non-linear dynamics
to the linear QNM ringdown regime.

Apart from the high SNR/precision requirement, QNM data analysis
may also have to face a ``confusion problem" when searching for
deviations from Kerr. This issue, already familiar from the modeling
of extreme mass ratio inspirals in non-Kerr
spacetimes~\cite{Glampedakis:2006sb}, has to do with the possibility
of misidentifying a true non-Kerr QNM signal with a Kerr one but with
a different set of mass and spin parameters. This degeneracy should
be broken by the simultaneous observation of the QNM frequency and
damping rate and/or the identification of more than one multipole
(see e.g.~\cite{Dreyer:2003bv, Berti:2005ys}).

As already emphasized, the backbone of our post-Kerr formalism
is the eikonal limit association between the spacetime's light
ring and the fundamental QNM. In the case of GR's garden-variety
black holes this connection is intuitively well-established, and
performs surprisingly well in approximating the rigorously computed
QNMs. Moreover, the fundamental mode is the one dominating the hole's
dynamical response in the time domain.

Since the GW150914 event, however, the light-ring/QNM connection
has been the subject of some debate. It has been shown, for instance,
that the connection is not as solid as one might think, since it is
in principle possible to construct spacetimes where the properties of
the wave potential are qualitatively different from those of the
geodesic potential for photons~\cite{Khanna:2016yow}. Indeed, a
spacetime may have no light rings and still exhibit a QNM ringdown
signal. Nevertheless, it should be pointed out that no such
counterexample has been constructed for black hole spacetimes resulting
from the field equations of a physically sensible modified gravity theory.

The light-ring/QNM connection has also been shown to fail in the
context of higher-dimensional Einstein-Lovelock black holes, as a
result of the perturbation equations having distinct eikonal limits
for different classes of gravitational perturbations~\cite{Konoplya:2017wot}
(in contrast, the connection has long been known to work for solutions of the
higher-dimensional Einstein equations, including Schwarzschild-Tangherlini and
Myers-Perry black holes~\cite{Cardoso:2008bp}). However, higher-dimensional
modifications of gravity are well constrained, and unlikely to give
measurable modifications in the context of testing the Kerr solution
in astrophysics~\cite{Kanti:2004nr,Berti:2015itd}. Furthermore, the
black hole counterexample constructed in~\cite{Konoplya:2017wot} is
known to exhibit instabilities at large values of the coupling
constant of the theory.

The upshot of this discussion is that, although both of the
aforementioned counterexamples on the light-ring/QNM connection are
interesting and instructive, they have little bearing on our post-Kerr
model, since they are not products of consistent modified gravity theories.
In the few cases where QNM calculations in such theories do exist (see
e.g.~\cite{Blazquez-Salcedo:2016enn}), the connection with the circular
photon orbit stands as firm as in GR.

Coming from the exactly opposite direction, a series of recent
papers~\cite{Cardoso:2016rao, Giudice:2016zpa, Cardoso:2016oxy, Mark:2017zmc}
uses the light-ring/QNM link to claim indistinguishability between
black holes and other horizonless compact objects.  Although these two
classes of systems are known to support markedly different QNM
spectra, they may indeed share the same QNM-like ringdown
signal\footnote{To our knowledge, this counterintuitive property was
  first noted by Nollert~\cite{Nollert:1996rf}, who replaced the
  standard Regge-Wheeler potential of Schwarzschild black holes with a
  potential made of a series of step functions. It had also been seen
  in the scattering of waves off the potential of compact relativistic
  stars (see e.g.~\cite{Kokkotas:1999bd, Ferrari:2000sr},
  but until recently this observation was largely overlooked.}.
This agreement, however, cannot persist for long, since horizonless
objects are expected to support a family of slowly damped $w$-modes in
the ``cavity" formed between the peak of the wave potential and the
body's center (or reflecting surface)~\cite{1991RSPSA.434..449C,Kokkotas:2003mh,Kokkotas:1994an}.
These modes should show up at a later stage of the signal, hence ending any
transient similarity with black hole dynamics~\cite{Barausse:2014tra}.
It should be noted that the degeneracy in the dynamical response of
these objects is partially due to the common exterior Schwarzschild
spacetime enforced by Birkhoff's theorem, so it is conceivable that
Kerr black holes may not always share the same ringing signal with
horizonless rotating bodies, simply because their light rings are
different.
The so-called ergoregion
instability~\cite{1978MNRAS.182...69S,1978RSPSA.364..211C,Cardoso:2007az}
(which sets in via the same trapped $w$-modes mentioned
earlier~\cite{Kokkotas:2002sf}) may provide another way of lifting
the degeneracy between these two types of systems.

As a final remark, it is worth mentioning that the notion of
non/post-Kerr light rings (albeit without their connection to QNMs)
has already been employed in the context of photon astronomy and the
models that are being developed as part of the ongoing effort to
produce direct images of our Galactic center supermassive black hole
(see e.g.~\cite{Johannsen:2015qca}). Although the basic methodology is
very different to that of gravitational wave astronomy, the two
efforts share the same ultimate goal of probing the physics of the
Kerr spacetime.

%%%%%%%%%%%%%%%%%%%%%%%%

\acknowledgements

E.B. was supported by NSF Grants No. PHY-1607130 and AST-1716715, and
by FCT contract IF/00797/2014/CP1214/CT0012 under the IF2014
Programme.
H.O.S. was supported by NSF Grant No. PHY-1607130. He also thanks Thomas
Sotiriou and the University of Nottingham for the hospitality in the final
stages of this work.
This work was supported by the H2020-MSCA-RISE-2015 Grant
No. StronGrHEP-690904.

%%%%%%%%%%%%%%%  APPENDIX %%%%%%%%%%%%%%%%%%%%%%%%%%%%%%%%%%
\appendix

\section{Fits of the offset function}
\label{app:offset}

\begin{figure}[htb]
\includegraphics[width=0.49\textwidth]{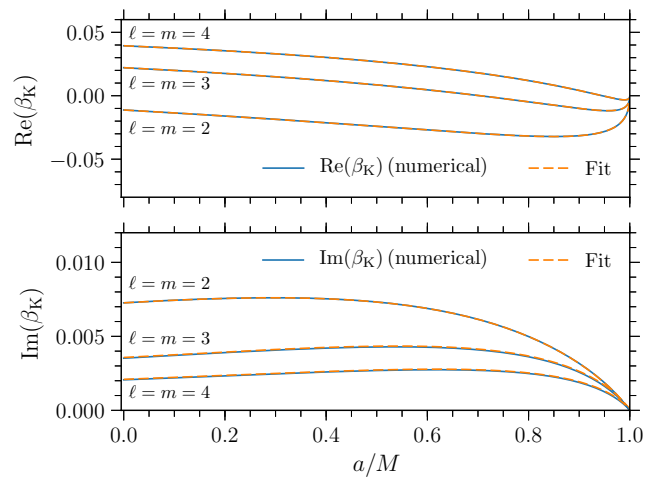}
\caption{
The real part (top panel) and imaginary part (bottom panel)
of the offset function $\beta_{\rm K}$ for $\ell = m = 2,\,3,\,4$.
The behavior is similar for higher values of $\ell = m$.  The
fitting coefficients $a_i$ for the real and imaginary parts of
$\beta_{\rm K}$ are listed in Table~\ref{tab:coeff}.
}
\label{fig:beta}
\end{figure}

In this appendix we present accurate fits for the offset function
$\beta_{\rm K}$. In Fig.~\ref{fig:beta} we show the behavior of
$\beta_{\rm K}$ as computed from Eq.~\eqref{eq:beta} for
modes with $\ell = m =2,\,3,\,4$.
We fitted $\beta_{\rm K}$ using the following function, inspired by
the classic interatomic Buckingham
potential~\cite{1938RSPSA.168..264B}:
\be
f(x) = a_1 + a_2 e^{-a_3 (1 - x)^{a_4}} - \frac{1}{a_5 + (1 - x)^{a_6}}.
\label{eq:buck}
\ee
The coefficients $a_i$ ($i = 1 \dots 6$) for the real and imaginary
parts of the leading $\ell=m$ offset functions (up to $\ell=m=7$) are
listed in Table~\ref{tab:coeff}.

As shown in Table~\ref{tab:coeff}, the absolute error (defined
as $\left\vert y_{\rm fit} - y_{\rm data}\right\vert$)
remains below $0.13$ in the interval $a/M \in [0, 0.9999]$. Our
results may not be reliable in the near extremal limit
($a/M \approx 1$), where the computation of QNMs in computationally
challenging and multipole QNM branches
exist~\cite{Yang:2012he,Yang:2013uba,Cook:2014cta,Richartz:2015saa,Richartz:2017qep}.
The QNM data tables (calculated using Leaver's continued fraction
method) used to obtain $\beta_{\rm K}$ are reliable
below the theoretical upper bound on the dimensionless spin of
astrophysical BHs (the so-called Thorne limit,
$a/M \approx 0.998$~\cite{1974ApJ...191..507T}, so our $\beta_{\rm K}$
fits should be adequate for astrophysical applications of the present
formalism.

%%%%%%%% FIG. 5 %%%%%%%%%%%%%%%%%%%%%%%

%%%%%%%% TABLE I %%%%%%%
\begin{table*}
\begin{tabular}{c c c c c c c c}
\hline
\hline
$\ell = m $  &  $a_1$  &  $a_2$  &  $a_3$ & $a_4$ & $a_5$ & $a_6$ &
{\rm max err.} \\
& & & & & & & $[10^{-2}]$ \\
\hline
2  &  0.1282(0.1381)  &  0.4178(0.3131)  &  0.6711(0.5531) & 0.5037(0.8492) & 1.8331(2.2159) & 0.7596(0.8544) & 0.023(0.004) \\

3  &  0.1801(0.1590)  &  0.5007(0.3706)  &  0.7064(0.6643) & 0.5704(0.6460) & 1.4690(1.8889) & 0.7302(0.6676) & 0.005(0.008) \\

4 & 0.1974(0.1575) & 0.4982(0.3478) & 0.6808(0.6577) & 0.5958(0.5840) &
1.4380(1.9799) & 0.7102(0.6032) & 0.011(0.009) \\

5 &  0.2083(0.1225)  &  0.4762(0.1993)  &  0.6524(0.4855) & 0.6167(0.6313) & 1.4615(3.1018) & 0.6937(0.6150) & 0.016(1.335) \\

6  &  0.2167(0.1280)  &  0.4458(0.1947)  &  0.6235(0.5081) & 0.6373(0.6556) & 1.5103(3.0960) & 0.6791(0.6434) & 0.021(0.665) \\

7  & 0.2234(-15.333)  &  0.4116(15.482)  &  0.5933(0.0011) & 0.6576(0.3347) & 1.5762(6.6258) & 0.6638(0.2974) & 0.025(0.874) \\

\hline
\hline
\end{tabular}

\caption{
The coefficients $a_{i}$ ($i = 1 \dots 6$) of the
fit~\eqref{eq:buck} for the offset function $\beta_{\rm K}$
[as defined in Eq.~\eqref{eq:beta}] for Kerr QNMs with
$\ell = m = 2 \dots 7$. Numbers outside (inside)
parentheses correspond to the real (imaginary) part of
$\beta_{\rm K}$, respectively. We also tabulate the largest
absolute error ($\equiv \left\vert y_{\rm fit}
- y_{\rm data}\right\vert$). The fits lose accuracy as we approach
the near extremal Kerr limit: we found that for all $\ell = m$
pairs the largest fitting error typically happens at
$a/M \approx 0.9999$.
}
\label{tab:coeff}
\end{table*}

%%%%%%%%%%%%%%%%%%%%%%%%%%%%%%%%%%%%%%

\section{Circular photon orbits in Kerr}
\label{sec:Kcirc}

This appendix provides a self-contained discussion of the properties of
the equatorial Kerr circular photon orbit.  Although this is well known
textbook material, we reproduce the calculation as a useful comparison for
the more general post-Kerr light ring analysis.

The equatorial Kerr metric in Boyer-Lindquist coordinates reads
\begin{align}
g_{tt}^\rK &= - \left (1-\frac{2M}{r} \right ), \quad g_{t\varphi}^\rK = -\frac{2Ma}{r},
\\
g_{\varphi\varphi}^\rK &= r^2 + a^2 + \frac{2Ma^2}{r}, \quad g_{rr}^\rK = \frac{r^2}{\Delta}.
\end{align}
Circular motion at the light ring radius $r_\rph$  simultaneously
solves $V_{\rm eff} (r_\rph)=0$ and $V^\prime_{\rm eff} (r_\rph) =0$.
We have,
\begin{align}
& r_\rph^3 + (a^2-b^2)r_\rph + 2M (a-b)^2 = 0,
\label{2conds_1}
\\
&r_\rph^3 - M (a-b)^2=0.
\label{2conds}
\end{align}
Elimination of $r_\rph^3$ leads to
\be
r_\rph = 3M \left ( \frac{b-a}{b+a} \right ) \quad \Leftrightarrow \quad
b = -a \left (\frac{r_\rph +3M}{r_\rph -3M} \right ).
\label{bformula}
\ee
These predict the correct radius $r_\rph = 3M$
for $a=0$, and also that prograde (retrograde) orbits should have
$r_\rph < 3M$ ($r_\rph > 3M$), but the $b(r_\rph)$ formula returns
an undetermined $0/0$ Schwarzschild limit.

Inserting $r_\rph (b)$ back into $V^\prime_{\rm eff} (r_\rph) = 0$
allows us to derive a cubic equation for the impact parameter:
\be
(a-b)^2 [\, 27 M^2 (a-b) + (a+b)^3 \, ]= 0.
\ee
The two real roots of this equation correspond to prograde and
retrograde motion.

A different (but completely equivalent) result $b(r_\rph)$
with a well-defined $a=0$ limit is given by the solution
of Eq.~(\ref{2conds}). This is
\be
b = a \pm \frac{r_\rph^{3/2}}{M^{1/2}} \equiv b_\rph,
\label{bKsol2}
\ee
where the upper (lower) sign corresponds to prograde (retrograde) motion.
Then, Eq.~(\ref{2conds_1}) becomes
\be
r_\rph^{3/2} -3M r_\rph^{1/2} \pm 2a M^{1/2} = 0,
\label{rphK}
\ee
which is the textbook formula solved by Eq.~(\ref{rph_book}).

The azimuthal orbital frequency at the light ring  is given
by the turning point formula (\ref{bOm}),
\be
\Omega_\rph = \frac{1}{b}.
\label{OmK1}
\ee
We can produce two equivalent expressions using either
(\ref{bformula}) or (\ref{bKsol2}). The former choice leads
to the result
\be
\Omega_\rph = \frac{3M -r_\rph}{a(r_\rph +3M)},
\label{OmK2}
\ee
with the inherited (and unattractive) property of a
$0/0$ Schwarzschild limit. On the other hand,  (\ref{bKsol2}) leads
to the textbook formula\footnote{The textbook approach is
that of Ref.~\cite{Bardeen:1972fi}: derive $E$, $L$ for circular
equatorial motion of a test particle; divide these to obtain the
impact parameter
$b = \pm M^{1/2} (\, r^2 \mp 2a M^{1/2} r^{1/2} + a^2 \,) /(\, r^{3/2} -2Mr^{1/2} \pm a M^{1/2} \, )$;
use this in $ \Omega = -(g_{tt}^\rK \,b + g_{t\varphi}^\rK )/ ( g_{t\varphi}^\rK
\,b + g_{\varphi\varphi}^\rK )$ to arrive at Eq.~(\ref{OmKbook}). Note
that Ref.~\cite{Bardeen:1972fi} skips the details of the
complicated calculation of $E$ and $L$, which is presented in
Chandrasekhar's book~\cite{Chandrasekhar:579245}.} :
\be
\Omega_\rph = \pm \frac{M^{1/2}}{r^{3/2}_\rph \pm a M^{1/2}}.
\label{OmKbook2}
\ee
Due to its well defined $a=0$ limit and its ``Keplerian" form, this
is the preferred formula for $\Omega_\rph$.

%%%%%%%%%%%%%%%%%%%%%%%%%%%%%%%%%%%%%%%%%

\section{Post-Kerr parameters}
\label{sec:coeffs}

In this appendix we list the various coefficients appearing in the post-Kerr analysis of the Lyapunov exponent.
Beginning with those appearing in $H_\rph$ [see Eq.~(\ref{Hph_eq})], we have
\begin{align}
M_{t\varphi} &= 2r_\rph^2 -3M r_\rph + a^2,
% \\
\nonumber \\
W_{t\varphi} & =  13 r^2_\rph -33M r_\rph + 8a^2,
% \\
\nonumber \\
K_{tt} & =   - (33 M^2 + a^2) r^2_\rph + 9  (9 M^2 -a^2) M r_\rph
\nonumber \\
& \quad  -\, 38 M^2 a^2,
% \\
\nonumber \\
Q_{tt} & =    -21 M r_\rph^2 + 2 (27 M^2 -a^2) r_\rph - 19 M a^2,
% \\
\nonumber \\
J_{tt} & =  \frac{1}{2}  \left [\,  15 M r_\rph^2 + ( a^2 -27 M^2) r_\rph + 11 M a^2  \, \right ].
\end{align}
Next, we list the coefficients of  the $h^{\pp}_{\mu\nu}, h^\p_{\mu\nu},
h_{\mu\nu}$ terms appearing in the expression for $N_\rph$
[Eq.~(\ref{Nph})]. For the $G$-coefficients we have:
%
% \begin{widetext}
\begin{align}
G_{\varphi\varphi} &=  - \left ( 351 M^6 + a^6 +787 M^4 a^2 + 157 M^2 a^4  \right)r^2_\rph \nonumber \\
&\quad +12 M \left (133 M^4 a^2 + 81 M^6 +3 a^4 M^2 -a^6 \right ) r_\rph
\nonumber \\
&  \quad - 48 M^2 a^2 \left (2 a^4 + 16 M^2 a^2 + 9 M^4 \right ),
% \\
\\
G_{tt} &=   - \left (10935 M^8 + 36666 M^6 a^2 +a^8 +15160 M^4 a^4 \right.
\nonumber \\
&\quad\left.+742 M^2 a^6 \right ) r^2_\rph
+ 6M \left ( 2769 M^4 a^4 -227 M^2 a^6  \right.
\nonumber \\
& \left. \quad  +\, 5103 M^8 + 13527 M^6 a^2 -4a^8 \right) r_\rph
\nonumber \\
&\quad-24 M^2 a^2 \left ( 485 M^2 a^4 + 567 M^6 +13 a^6 \right.
\nonumber \\
&\left.\quad+1581 M^4 a^2 \right ),
% \\
\\
G_{t\varphi} &=   -3 M \left ( a^2 + 27 M^2 \right ) \left (83 a^4 + 262 M^2 a^2 + 87 M^4 \right) r^2_\rph
\nonumber \\
&\quad+ \left (19683 M^8 -727 M^2 a^6 + 46683 M^6 a^2 \right.
\nonumber \\
& \left. \quad   +\, 6941 M^4 a^4 -4 a^8 \right) r_\rph
\nonumber \\
&\quad-12 M a^2 \left (7 a^6 + 459 M^2 a^4 + 729 M^6 + 1829 M^4 a^2 \right ).
\nonumber \\
\end{align}
For the $Z$-coefficients the expression are:
\begin{align}
Z_{tt} &=    -M \left (  131a^6+6345M^4a^2+3379a^4 M^2\right.
\nonumber \\
&\quad \left. + 729M^6 \right )r^2_\rph
% \nonumber \\
% &\quad
+  \left ( 2187 M^8 -2 a^8 -361 M^2 a^6 \right.
%  \right.
% \nonumber \\
% & \left.  \quad +\,
\nonumber \\
&\quad \left.
+15309 M^6 a^2 + 4035 M^4 a^4   \right )r_\rph
\nonumber \\
&\quad - 4 M a^2 \left (  11 a^6 +1737 M^4 a^2 +243 M^6 +655 a^4 M^2  \right ),
\\
% \\
Z_{\varphi\varphi} &=  -M \left ( 154M^2a^2 +35 a^4 +27 M^4  \right ) r^2_\rph
\nonumber \\
&\quad+ \left ( 81M^6 +348 M^4 a^2 +5a^4 M^2  -2a^6  \right ) r_\rph
\nonumber \\
&   \quad -\, 4 M a^2 \left (  5a^4 +9 M^4 +40M^2 a^2  \right ),
\\
% \nonumber \\
Z_{t\varphi} & = - \left ( 1917M^4 a^2 + 845 a^4 M^2 + 19a^6 +243M^6  \right )r^2_\rph
\nonumber \\
&\quad + M \left (   847a^4 M^2 -91 a^6 +729M^6
  % \right.
% \nonumber \\
% & \left.  \quad
+\, 4563M^4 a^2 \right ) r_\rph
\nonumber \\
&\quad - 4a^2 \left ( a^6 +81M^6 +519M^4 a^2 +155 a^4 M^2 \right),
\end{align}
The $E$-coefficients are given by:
\begin{align}
 E_{rr} &=  \left (4 a^6 +154 M^2 a^4  +135M^6 +436 M^4 a^2 \right ) r^2_\rph
 \nonumber \\
 &\quad- 2M \left (189 M^6 -8a^6 +70  M^2 a^4 + 478 M^4 a^2 \right ) r_\rph
\nonumber \\
&   \quad +\, a^2 \left  (112 M^2 a^4  + 448 M^4 a^2 + a^6 +168 M^6 \right ),
\\
% \nonumber \\
E_{tt} & =    - \left (  549 M^4 a^2 -a^6 +1377 M^6 -161a^4 M^2  \right ) r^2_\rph
\nonumber \\
&\quad + 3M \left ( 15M^2 -a^2\right ) \left ( 8 M^2 a^2 +81M^4 -5a^4 \right ) r_\rph
\nonumber \\
&\quad   -4M^2 a^2 \left ( 405M^4 -29 a^4 +65M^2 a^2 \right ),
\\
% \nonumber \\
E_{\varphi\varphi} &=  - \left (39 M^4   -a^4 -2 M^2 a^2 \right  ) r^2_\rph
\nonumber \\
&\quad+ 3 M \left ( 33M^4 +a^4 -10M^2 a^2 \right ) r_\rph
\nonumber \\
&\quad -4 M^2 a^2 \left  (  11M^2-2a^2 \right ),
\\
% \nonumber \\
E_{t\varphi} &=  - 4M \left ( 54M^4 +14M^2 a^2 - 5a^4  \right )  r^2_\rph
\nonumber \\
&\quad + \left  ( 567 M^6 -45 M^4 a^2 -19 a^4M^2 +a^6 \right) r_\rph
\nonumber \\
& \quad -\, 12 M a^2 ( 21M^4 +M^2 a^2 -a^4 ).
\end{align}
Finally, the $S$-coefficients are:
\begin{align}
S_{tt} &=   9 \left (  3673M^4a^2 +1053 M^6 +947 a^4 M^2 +11 a^6 \right ) r^2_\rph
\nonumber \\
 &\quad -  M \left ( 4151 a^4 M^2 +71901 M^4 a^2 +26973 M^6  \right.
 \nonumber \\
 &\quad \left.-713a^6 \right ) r_\rph
% \nonumber \\
% &\quad
+ 4a^2 \left (  8409 M^4 a^2 +2997 M^6 \right.
\nonumber\\
&\quad \left.+1379 a^4 M^2 +4 a^6 \right ),
\\
% \nonumber \\
S_{\varphi\varphi} &= 3 ( 226M^2 a^2 +105 M^4 +17 a^4) r^2_\rph
\nonumber \\
&\quad - M ( 1298M^2 a^2 +891 M^4 - 101 a^4 ) r_\rph
\nonumber \\
&\quad + 4a^2 ( 158M^2 a^2 +99 M^4 +4 a^4 ),
\\
% \nonumber \\
S_{t\varphi} &=  (r_\rph +3M)(Mr_\rph)^{-1/2}
\nonumber \\
&\quad \times\left [\, M ( 2770M^2 a^2 +999 M^4 +407 a^4   ) r_\rph^2  \right.
\nonumber \\
& \left. \quad -\, (  2835 M^6 +5706 M^4 a^2 -173a^4 M^2  -16 a^6) r_\rph \right.
\nonumber \\
&\left.\quad + 12 Ma^2 ( 226 M^2 a^2 +105 M^4 +17 a^4 )  \, \right ].
\end{align}

%%%%%%%%%%%%%%%%%%%%%%%%%%%%%%%%%%%%%%%
\section{The Johannsen-Psaltis expansion coefficients}
\label{app:JPcoeffs}

Here we list the coefficients appearing in the expansions in
terms of $\varepsilon_3$ of the radius of the photon orbit, the
impact parameter, and the Lyapunov exponent for the JP spacetime.

First we define the auxiliary coefficients,
$(a+b_\rph) \equiv C_+$ and $(a-b_\rph) \equiv C_-$, which scale
as the mass and the auxiliary coefficient
$\left(27 M^2 C_- (4 a+b_\rph)+2 C_+^4\right) \equiv C_0$, which
scales as the mass to the fourth power.

Taking these definitions into account, the various coefficients
have the form
\begin{widetext}
\bear      \delta r_1 &=&   -\frac{b_\rph M C_+^5}{18 C_-^2 C_0},  \\
              \delta r_2 &=& \frac{M C_+^8}{972 C_-^5 C_0^3} \left[2916 M^4 C_-^2 \left(15 a^3-14 a^2 b_\rph-a b_\rph^2+b_\rph^3\right)%\nonumber\\
%              &&
              +27 M^2 C_- \left(60 a^3-16 a^2 b_\rph-33 a b_\rph^2+b_\rph^3\right) C_+^3 \right.\nonumber\\
              &&\left.+\left(15 a^3+6 a^2 b_\rph-11 a b_\rph^2-5 b_\rph^3\right) C_+^6\right],\\
                 \delta r_3 &=& \frac{M C_+^{11}}{52488 C_-^8
                 C_0^5} \left[4251528 M^8 C_-^4 \left(545 a^5-636 a^4 b_\rph-140 a^3 b_\rph^2+80 a^2 b_\rph^3+9 a
   b_\rph^4-2 b_\rph^5\right) \right.\nonumber\\
   &&+19683 M^6 \left(C_+ C_-\right)^3 \left(9020 a^5-6992 a^4 b_\rph-6614 a^3 b_\rph^2+416 a^2 b_\rph^3+526 a b_\rph^4+31
   b_\rph^5\right)\nonumber \\
   &&+729 M^4 C_-^2 \left(6990 a^5-2296 a^4 b_\rph-8075 a^3 b_\rph^2-1628 a^2 b_\rph^3+599 a b_\rph^4+132 b_\rph^5\right)
   C_+^6 \nonumber\\
   &&+27 M^2 C_- \left(2405 a^5+404 a^4 b_\rph-3412 a^3 b_\rph^2-1836 a^2 b_\rph^3+101 a b_\rph^4+106
   b_\rph^5\right) C_+^9\nonumber\\
   &&\left.+2 \left(155 a^5+110 a^4 b_\rph-222 a^3 b_\rph^2-236 a^2 b_\rph^3-37 a b_\rph^4+13 b_\rph^5\right)
   C_+^{12}\right],\\
%   \eear
% \bear
   \delta b_1 &=& \frac{54 M^2 C_- C_+{}^4+C_+{}^7}{54 C_-{}^2 C_0},\\
   \delta b_2 &=& \frac{C_+^7}{1944 C_-^5 C_0^3} \left[78732 M^6 C_-^3
    \left(29 a^2+4 a b_\rph-b_\rph^2\right)+729 M^4 C_-^2
   \left(204 a^2+88 a b_\rph+b_\rph^2\right) C_+^3\right.\nonumber\\
   &&+27 M^2 C_- \left(117 a^2+96 a b_\rph+13 b_\rph^2\right)
   C_+^6\left.+2 \left(11 a^2+14 a b_\rph+4 b_\rph^2\right) C_+^9\right],  \\
   \delta b_3 &=& \frac{C_+^{10}}{314928 C_-^8 C_0^5} \left[114791256 M^{10} C_-^5 \left(3737 a^4+938 a^3 b_\rph-258 a^2
   b_\rph^2-46 a b_\rph^3+5 b_\rph^4\right)\right.\nonumber\\
   &&+1062882 M^8 C_-^4 \left(43430 a^4+22268 a^3
   b_\rph-195 a^2 b_\rph^2-1150 a b_\rph^3-85 b_\rph^4\right) C_+^3\nonumber\\
   && +19683 M^6
   C_-^3 \left(99430 a^4+81868 a^3 b_\rph+11013 a^2 b_\rph^2-3450 a b_\rph^3-675
   b_\rph^4\right) C_+^6 \nonumber\\
   &&  +729 M^4 C_-^2 \left(56225 a^4+66218 a^3 b_\rph+17502 a^2
   b_\rph^2-2002 a b_\rph^3-805 b_\rph^4\right) C_+^9 \nonumber\\
   && +54 M^2 C_- \left(7870
   a^4+12370 a^3 b_\rph+5043 a^2 b_\rph^2-182 a b_\rph^3-251 b_\rph^4\right) C_+^{12} \nonumber\\
   && \left.+4
   \left(437 a^4+875 a^3 b_\rph+501 a^2 b_\rph^2+15 a b_\rph^3-39 b_\rph^4\right)
   C_+^{15}\right],\\
% \eear
% \bear
 \delta\gamma_1&=&\gamma _{\text{ph}}^3  \frac{M^4\left(27 M^2 C_-^2
     +a \left(a-2 b_{\text{ph}}\right) C_+^2\right) }{2
   C_+^5C_0 \left(3 M^2 \left(5 a - b_{\text{ph}}\right) C_- + a^2 C_+^2\right)^3}\left[a^3 \left(4 a^2-a b_{\text{ph}}-6 b_{\text{ph}}^2\right) C_+^{10} \right.\nonumber\\
   &&+729 M^6
   C_-^3 \left(364 a^4-227 a^3 b_{\text{ph}}-201 a^2 b_{\text{ph}}^2-29 a b_{\text{ph}}^3+13 b_{\text{ph}}^4\right)
   C_+^2\nonumber\\
   && +27 M^4 C_-^2 \left(2 a+b_{\text{ph}}\right) \left(319 a^4-174 a^3 b_{\text{ph}}-216 a^2 b_{\text{ph}}^2-38 a
   b_{\text{ph}}^3+9 b_{\text{ph}}^4\right) C_+^4 \nonumber\\
   && \left.+a M^2 C_- \left(454 a^4-133 a^3 b_{\text{ph}}-366 a^2
   b_{\text{ph}}^2-182 a b_{\text{ph}}^3+2 b_{\text{ph}}^4\right)
   C_+^7+78732 M^8 C_-^5 \left(5 a-b_{\text{ph}}\right) \left(4 a+b_{\text{ph}}\right)\right].
    \eear
\end{widetext}

\end{document}